\documentclass[twocolumn]{aastex631}

\newcommand{\msun}{M_{\odot}}
\newcommand{\mstar}{M_{\star}}
\newcommand{\mhalo}{M_{\rm halo}}
\newcommand{\rhalo}{r_{\rm halo}}
\newcommand{\mvir}{M_{\rm vir}}
\newcommand{\rvir}{r_{\rm vir}}
\newcommand{\rhalf}{r_{\star}^{50}}

\usepackage{hyperref}
\usepackage{amsmath}
\usepackage{booktabs}

\usepackage[table]{xcolor}
\definecolor{BrickRed}{RGB}{182,50,28}

\usepackage{array,booktabs,colortbl}
\newcolumntype{P}[1]{>{\raggedright\arraybackslash}p{#1}}

\begin{document}

\title{Effects of Environment on the Star Formation Histories and Shapes of Simulated Galaxies}

\author[0000-0002-5908-737X]{Francisco J. Mercado}
\affiliation{Department of Physics and Astronomy, University of Southern California, Los Angeles, CA 90089, USA}
\affiliation{Department of Physics \& Astronomy, Pomona College, Claremont, CA 91711, USA}
\email{francisco.mercado@usc.edu}

\author[0009-0008-9144-1747]{Brian A. Giraud Calderon}
\affiliation{Department of Physics \& Astronomy, Pomona College, Claremont, CA 91711, USA}
\affiliation{Department of Astronomy, Wesleyan University, Middletown, CT 06459, USA}

\author[0000-0002-3430-3232]{Jorge Moreno}
\affiliation{Department of Physics \& Astronomy, Pomona College, Claremont, CA 91711, USA}
\affiliation{Carnegie Observatories, 813 Santa Barbara st, Pasadena, CA 91101, USA}

\author[0000-0002-2651-7281]{Coral Wheeler}
\affiliation{Department of Physics \& Astronomy, California Polytechnic State University Pomona, Pomona, CA 91768, USA}

\author[0000-0003-1896-0424]{Jose A. Benavides}
\affiliation{Department of Physics and Astronomy, University of California, Riverside, 900 University Avenue, Riverside, CA 92521, USA}

\author[0000-0003-0965-605X]{Pratik J. Gandhi}
\affiliation{Department of Astronomy, Yale University, New Haven, CT 06520, USA}

\author[0000-0002-1732-8040]{Tyler Smith}
\affiliation{Department of Physics and Astronomy, University of Southern California, Los Angeles, CA 90089, USA}

\author[0009-0002-3536-1220]{Olti Myrtaj}
\affiliation{Department of Physics and Astronomy, University of California, Irvine, CA 92697, USA}

\author[0000-0002-1109-1919]{Robert Feldmann}
\affiliation{Department of Astrophysics, University of Zurich, Winterthurerstrasse 190, CH-8057 Zurich, Switzerland}

\begin{abstract}
We investigate how environment is connected to the formation times and three-dimensional stellar shapes of galaxies using the FIREbox cosmological volume simulation. We analyze 1720 resolved galaxies at $z=0$ spanning stellar masses $\mstar\approx5\times10^6$--$5\times10^{11}\,M_\odot$ and environments ranging from isolated galaxies to groups in $\sim10^{13}\,M_\odot$ halos. We quantify environment using the Perturbation Index (PI), which measures the cumulative tidal influence of neighboring galaxies relative to each galaxy's own binding force. We measure formation times from cumulative star formation histories and shapes from intrinsic stellar axis ratios. Formation times depend most strongly on stellar mass, but below $10^{10}\,M_\odot$, galaxies in more strongly perturbing environments formed their stars earlier, most clearly in the late-time star formation of low-mass galaxies. Galaxies in these environments also have larger minor-to-major axis ratios at all masses: more spherical at low and intermediate mass and thicker along the minor axis at high mass. A partial-correlation analysis shows that, below $10^8\,M_\odot$, the environment--shape correlation is no longer significant at fixed formation time, whereas between $10^8$ and $10^{10}\,M_\odot$ a modest correlation remains. Above $10^{10}\,M_\odot$, the weak environment--shape correlation is not explained by formation time. Below $10^{10}\,M_\odot$, high-PI galaxies also have lower halo masses at fixed stellar mass; at fixed halo mass, environment is only weakly correlated with formation time and shape, whereas formation time and shape remain correlated.
We therefore predict that, at fixed stellar mass, dwarf galaxies in stronger tidal environments should appear rounder in projection, with the difference largely tracking the ages of their stellar populations.
\end{abstract}

\keywords{Galaxy evolution (594) -- Galaxy formation (595) -- Dwarf galaxies (416) -- Galaxy structure (622) -- Galaxy interactions (600) -- Star formation (1569)}

\section{Introduction}\label{sec:intro}  
Galaxy evolution reflects the interplay between processes internal to galaxies and the larger-scale environments in which they reside \citep{Mo2010,Somerville2015,Naab2017}. Although galaxy growth depends most strongly on stellar mass, galaxies do not evolve in isolation. Their surroundings can influence when and how they form stars, particularly in dense regions where interactions with the larger-scale environment become important \citep{Boselli2006,Blanton2009,Peng2010,Guglielmo2015,GarrisonKimmel2019b,Wang2024}. Disentangling internal and environmental effects remains a critical challenge because stellar mass, environment, star formation history (SFH), and morphology are all coupled. Galaxies in dense environments are therefore not simply field galaxies that have experienced additional environmental processing. Instead, they may follow intrinsically different evolutionary pathways because they form within halos with different assembly histories \citep{Guglielmo2015,GarrisonKimmel2019b,Digby2019,MonteroDorta2021}. This possibility is related to halo assembly bias, in which the clustering of halos of fixed mass depends on their assembly history \citep{Gao2005,Wechsler2006}. Understanding how processes internal to galaxies combine with the external environment to shape galaxy evolution remains one of the core goals of modern galaxy formation studies.

Large observational surveys illustrate why stellar mass must be treated carefully when interpreting environmental trends. \citet{Kauffmann2004} show that several structural and SFH indicators depend strongly on stellar mass, while environmental trends become most apparent in dense regions and on local scales. More recently, \citet{Bluck2019} frame the relationship between galaxy properties and structure as a predictor-ranking problem, finding that stellar mass and distance from the star-forming main sequence are more predictive of structure, as measured by bulge-to-total ratio, than multiple commonly used environmental metrics. These results demonstrate that environmental trends should not be interpreted independently of stellar mass. Differences between galaxies in overdense and underdense environments may partly reflect differences in the stellar mass functions of those environments, while genuine environmental signatures can be diluted or hidden if the dominant dependence on stellar mass is not controlled for explicitly.

Despite this strong mass dependence, environment influences galaxy evolution beyond the connection with stellar mass. At fixed stellar mass, galaxies in different environments exhibit differences in their star formation histories and present-day structural properties. \citet{Guglielmo2015} examine SFHs as a function of environment, stellar mass, and morphology, finding that cluster galaxies have accelerated formation histories compared with field galaxies even when controlling for present-day mass and morphology. Similarly, \citet{PerezMillan2023} emphasize that local environment measures and morphology must be considered together in order to identify the mechanisms through which environments influence galaxy evolution. These studies establish that environmental effects on galaxy growth cannot be fully reduced to stellar mass or morphology alone. This motivates approaches that treat environment, mass, and star formation history as interconnected quantities.

The relationship between environment and galaxy structure is similarly complex because morphology reflects both the underlying stellar distribution and the recent star formation state of a galaxy. Large galaxy surveys show that environmental trends in color and star formation are often stronger than environmental trends in visual morphology at fixed stellar mass. For example, Galaxy Zoo studies find that the morphology-density relation is substantially weaker than the color-density relation once stellar mass is accounted for \citep{Bamford2009}. \citet{vanderWel2008} further highlight the distinction between galaxy \textit{structure} and visually defined \textit{morphology}, showing that structural parameters, such as the S\'ersic index, primarily depend on stellar mass while visual morphology can be strongly influenced by star formation and stellar populations. This distinction is important because broad morphological classifications can obscure continuous changes in the underlying stellar distribution from galaxy to galaxy. Therefore, a galaxy population might undergo significant structural evolution without transitioning cleanly between traditional morphological classes.

These limitations are particularly relevant for low-mass galaxies, where environment, star formation history, and structure are more strongly connected. While the evolution of massive galaxies is often strongly regulated by internal processes associated with their deeper gravitational potentials and self-regulated star formation, low-mass galaxies are more susceptible to environmental perturbations because their shallower potentials make them more vulnerable to gas loss and changes in their baryon supply \citep{Dekel1986, Mayer2001,Kauffmann2004,Mayer2006,Peng2010,Somerville2015,Simpson2018,Digby2019,Weisz2011,Weisz2014}. Resolved stellar population studies of low-mass galaxies in the Local Group and Local Volume reveal strong connections between environment and morphology, with dwarf spheroidals (dSphs) preferentially found in denser regions or as satellites of more massive hosts \citep{Goto2003,McConnachie2012}. \citet{Weisz2011} show that dSphs in the ACS Nearby Galaxy Survey Treasury (ANGST) sample are less isolated than dwarf irregulars (dIrrs) and argue that external mechanisms, including tidal effects and ram-pressure stripping, are likely required to explain the transition from gas-rich to gas-poor systems. Their analysis of the SFHs of low-mass Local Group galaxies further demonstrates that the fraction of stellar mass formed at early times varies strongly among low-mass galaxies and is closely connected to environmental differences \citep{Weisz2014}. Together, these observations suggest that low-mass galaxies provide an important laboratory for understanding how environment modifies both star formation histories and galaxy structure.

Cosmological simulations provide an additional framework for connecting environmental histories to the physical mechanisms that drive galaxy evolution. Simulations from the Feedback In Realistic Environments (FIRE) project\footnote{\url{https://fire.northwestern.edu/}} using the FIRE-2 model \citep{Hopkins2018} demonstrate that environmental effects are not fully captured by a simple satellite versus central classification, as galaxies can experience environmental influences before or without becoming satellites of massive hosts. \citet{GarrisonKimmel2019b} show that low-mass central galaxies in the environments of Milky Way-mass hosts form their stars earlier than comparable isolated galaxies, with environmental effects extending beyond the virial radii of massive hosts and affecting both satellites and centrals residing in Local Group-like regions. They also find that more massive galaxies form their stars later regardless of environment, highlighting the importance of stellar mass as a primary driver of galaxy star formation histories. The APOSTLE \citep{Sawala2016} and Auriga \citep{Grand2017} simulations of Local Group-like environments similarly find that low-mass satellites have suppressed recent star formation relative to comparable field galaxies, likely due to environmental removal of gas after infall \citep{Digby2019,Simpson2018}. Other cosmological low-mass galaxy simulations extend this picture by showing that environmental effects can alter both star formation histories and structural evolution \citep[e.g.,][]{Christensen2024}. Using FIREbox, \citet{Mercado2025} found that, at fixed stellar mass, low-mass galaxies in stronger tidal environments are generally larger and have lower dark matter content than those in weaker tidal environments. In the TNG50 simulation, the cumulative SFHs of dwarf galaxies depend mainly on stellar mass and on whether a galaxy is a central or a satellite, with satellites assembling their stellar mass earlier \citep{Joshi2021}, and in FIRE-2 simulations, most satellites below $\mstar\sim10^7\,\msun$ around Milky Way-mass hosts are quiescent \citep{Samuel2022}. Stellar masses also depend on halo assembly history: at fixed halo mass, low-mass central galaxies in FIRE simulations at $z=2$--6 have lower stellar masses in faster-growing halos and higher stellar masses in more concentrated, earlier-forming halos \citep{Feldmann2019}. Together, these simulations demonstrate that environment can influence when low-mass galaxies form their stars and how they evolve over time. This connection between star formation history and structure is physically motivated by models of environmental transformation such as tidal stirring, in which repeated tidal encounters can suppress star formation and dynamically heat stellar orbits \citep{Mayer2001,Kazantzidis2011,Lokas2012,Lokas2015}. Environment also leaves an imprint on the structure of the dark matter halos that host galaxies: in cosmological zoom-in simulations of Milky Way-mass halos, \citet{Arora2025} find that the halo's quadrupole distortion, which traces its triaxial shape, aligns with the largest surrounding filaments and persists into the inner $\sim30$\,kpc, while infalling satellites produce more transient distortions.

The above studies therefore establish important connections between environment, stellar mass, SFH, and morphology, but these relationships are often examined separately. Many observational studies rely on instantaneous star formation rates, colors, or quenching indicators rather than cumulative star formation histories. Because cumulative SFHs record when a galaxy's present-day stars formed over its entire lifetime, they can retain the imprint of past environmental effects that a present-day star formation rate or color does not capture. Furthermore, morphology is frequently characterized through broad visual classifications, which may not capture continuous changes to the underlying stellar distribution. Additionally, environmental effects are often quantified using discrete categories such as field, group, or cluster membership, which can obscure the continuous nature of tidal perturbations. As a result, it remains unclear whether environmental trends in galaxy structure arise primarily because galaxies in more strongly perturbing environments form their stars earlier, because environment directly alters stellar dynamics, because both reflect differences in halo assembly, or through a combination of these effects.

\begin{figure*}
    \centering
    \includegraphics[width=1.0\textwidth, trim = 0 0 0 0]{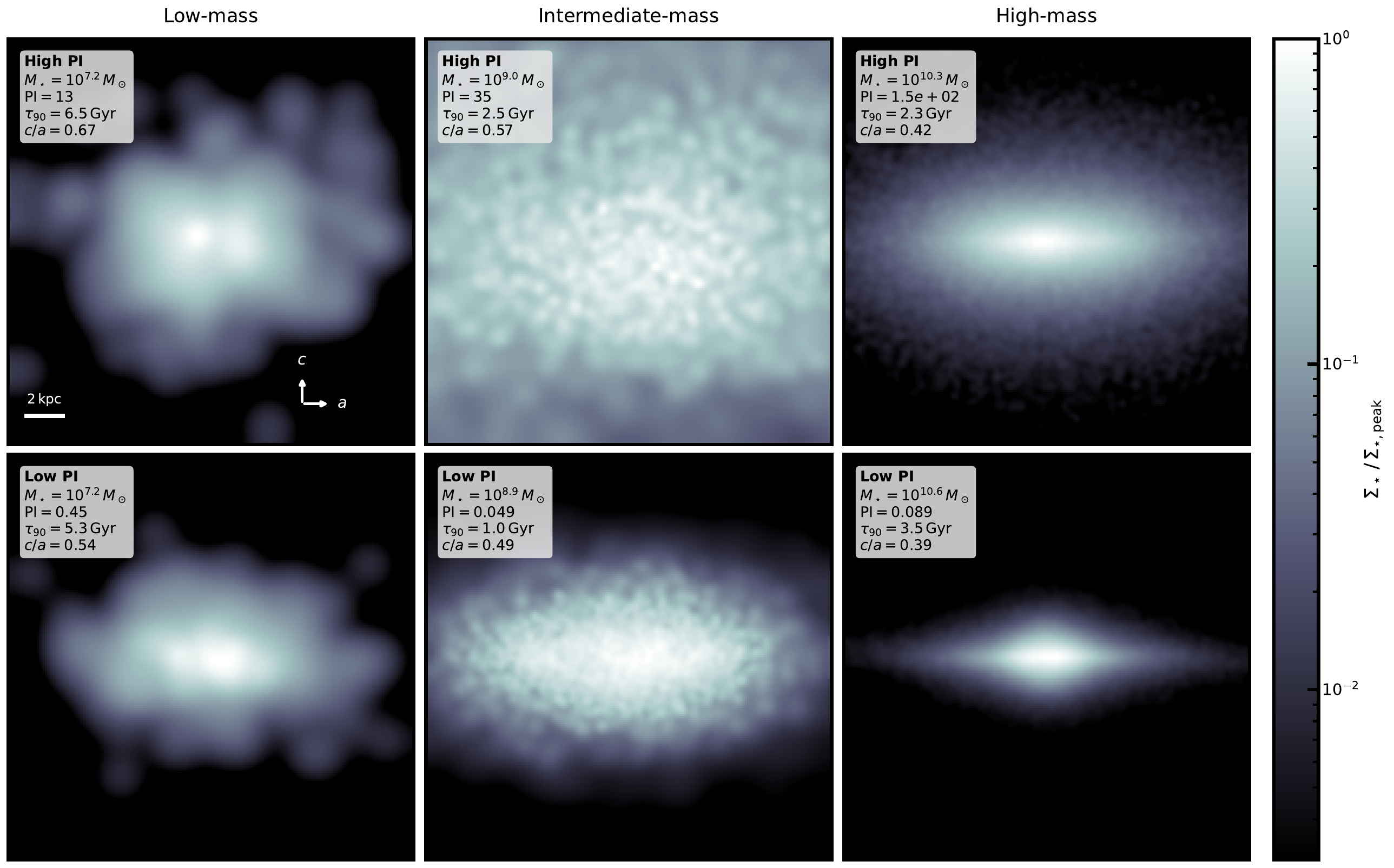}
    \caption{Example FIREbox galaxies illustrating the main quantities used in this work. \textbf{Columns:} galaxies near the median stellar mass of each of the three mass bins (Section~\ref{sec:sample_selection}). \textbf{Top row:} high-PI (${\rm PI}>1$) galaxies; \textbf{bottom row:} low-PI (${\rm PI}<1$) galaxies at similar stellar mass.
    Each panel shows the peak-normalized, mass-weighted stellar surface density, centered on the stellar center and oriented into the galaxy's major/minor-axis ($a$, $c$) frame from the shape fit (Section~\ref{sec:shape_measurements}).}
    \label{fig:intro_gallery}
\end{figure*}

In this work, we address this question using the FIREbox cosmological volume simulation \citep{Feldmann2023}. FIREbox provides a large sample of resolved galaxies spanning a broad range of environments while maintaining the resolution necessary to measure both cumulative SFHs and intrinsic stellar structure. We quantify environment using a continuous Perturbation Index (PI), which estimates the cumulative strength of tidal perturbations from neighboring galaxies rather than assigning galaxies to discrete environmental categories. We quantify when galaxies formed their stars using $\tau_{30}$, $\tau_{50}$, and $\tau_{90}$, corresponding to the lookback times by which galaxies formed 30, 50, and 90 percent of their present-day stellar mass. Finally, we quantify stellar structure using continuous three-dimensional axis ratios, focusing particularly on the minor-to-major axis ratio ($c/a$) as a measure of the thickness and roundness of stellar distributions. Figure~\ref{fig:intro_gallery} previews these quantities together on six example FIREbox galaxies before we define them formally in Section~\ref{sec:methodology}.

Our goal is to determine how environment, formation time, and galaxy structure are connected, and how these connections change with stellar mass, within a single cosmological population spanning five decades in stellar mass. Specifically, we ask three core questions: 1) How do galaxy formation times vary with present-day tidal environment in bins of stellar mass? 2) Do galaxy shapes exhibit continuous environmental trends in bins of stellar mass? 3) Does formation time statistically account for the connection between environment and shape, or does environment remain correlated with stellar structure after controlling for both formation history and stellar mass? By addressing these questions simultaneously, this work connects several previously separate lines of investigation into galaxy environment, SFHs, and morphology, providing a more complete picture of how galaxies respond to their surroundings.

The paper is organized as follows. In Section~\ref{sec:methodology} we describe the FIREbox simulation, galaxy sample, measurements of SFHs and stellar shapes, the PI, and our partial correlation analysis. In Section~\ref{sec:results}, we present environmental trends in formation times, stellar structure, and the interplay between these two quantities, and we examine how these trends depend on halo mass. In Section~\ref{sec:discussion}, we place the formation-time trends in context, discuss physical interpretations of the connection between environment and shape, and describe the caveats of our analysis. Finally, we summarize our conclusions in Section~\ref{sec:conclusion}.

\section{Methodology}\label{sec:methodology}

\subsection{The FIRE-2 galaxy formation model}\label{sec:FIRE2_model}

This work uses the FIREbox cosmological volume simulation \citep{Feldmann2023}, which adopts the FIRE-2 galaxy formation model \citep{Hopkins2018}. FIREbox is run with the \textsc{GIZMO} code \citep{Hopkins2015} using the meshless finite-mass (MFM) hydrodynamics method. The model follows the coupled evolution of dark matter, gas, and stars in a cosmological setting, while explicitly modeling the interstellar-medium physics and stellar feedback processes that regulate galaxy formation.

FIRE-2 follows radiative heating and cooling of gas over temperatures of $10$--$10^{10}\,{\rm K}$. Rather than imposing an effective equation of state on the interstellar medium (ISM), FIRE-2 explicitly models its dense, clumpy, multiphase structure. Star formation is restricted to gas that is dense, molecular, self-shielding, locally self-gravitating, and Jeans unstable. In FIREbox, stars can only form above a density threshold of $n = 300\,{\rm cm^{-3}}$. Once stars form, feedback is modeled through the injection of energy, momentum, mass, and metals from Type Ia and core-collapse supernovae, stellar winds from massive stars and asymptotic giant branch stars, radiation pressure, photo-ionization heating, and photo-electric heating. These feedback channels are implemented using stellar population properties from \textsc{Starburst99} \citep{Leitherer1999}, assuming a Kroupa initial mass function \citep{Kroupa2002}. Although each star particle represents a single stellar population that samples this initial mass function, FIRE-2 treats supernovae as individual, time-resolved events from each star particle, rather than injecting the combined energy of an IMF-averaged population at once. The simulations track 11 elements (H, He, and nine metals) and include a spatially uniform, redshift-dependent ultraviolet and X-ray background \citep{FaucherGiguere2009}.

Cosmological zoom-in simulations have benchmarked the FIRE-2 model against a wide range of observations. Simulations of low-mass galaxies in Local Group-like environments produce stellar mass functions, radial distributions, elemental abundances, diversity of star formation histories, and a stellar mass--halo mass relation comparable to those observed in the Local Group \citep{Wetzel2016,Escala2018,GarrisonKimmel2019a,GarrisonKimmel2019b,Samuel2020,Santistevan2023}. Simulations of Milky Way-mass galaxies agree with observed properties of the Milky Way, including giant molecular cloud lifetimes, stellar dynamics and kinematics, and elemental abundance patterns \citep{Benincasa2020,Bellardini2021,Bellardini2022,McCluskey2025,Graf2025a,Graf2025b,Bellardini2026,Myrtaj2026}. FIRE-2 galaxies spanning ultra-faint to Milky Way masses also agree well with the observed star-forming main sequence and stellar mass--metallicity relation at $z=0$ \citep{Gandhi2022}. Although these comparisons use zoom-in simulations rather than FIREbox, they show that the FIRE-2 model used in FIREbox reproduces these emergent galaxy properties.

\subsection{The FIREbox cosmological volume simulation}\label{sec:FIREbox}

FIREbox \citep{Feldmann2023} is a periodic cosmological simulation evolved from $z=120$ to $z=0$ in a box with side length $22.1\,{\rm cMpc}$. Its initial conditions were generated with \textsc{MUSIC} \citep{Hahn2011} assuming a flat $\Lambda$CDM cosmology consistent with \citet{PlanckCollaboration2016}. The flagship \texttt{FB1024} realization contains $1024^3$ gas resolution elements and $1024^3$ dark matter particles. The baryonic and dark matter particle masses are $m_{\rm b}=6.3\times10^4\, \msun$ and $m_{\rm DM}=3.3\times10^5\, \msun$, respectively. Gas gravitational softenings are adaptive and reach a minimum of 1.5 pc in the densest gas, while star particles and dark matter particles have fixed gravitational softenings of 12 pc and 80 pc. A more representative measure of the spatial resolution in the ISM is the inter-particle spacing of gas eligible for star formation, which is $\lesssim20$\,pc in FIREbox \citep{Feldmann2023}.

The combination of cosmological volume and high mass resolution makes FIREbox particularly useful for studying environmental effects across a wide range of galaxy masses. While its volume is smaller than that of many large cosmological simulations \citep[e.g.,][]{Schaye2015, Pillepich2018b}, FIREbox resolves scales from the interstellar medium to the larger-scale galaxy environment. At $z=0$, the simulation contains galaxies spanning a broad range of $\mstar$ and environments, from isolated field systems to galaxies in small groups hosted by halos approaching $10^{13}\, \msun$.

We use the $z=0$ snapshot of FIREbox, which has already been used to investigate environmental effects on low-mass galaxy structure and scaling relations \citep{Mercado2025,Benavides2025,Moreno2025,Klein2026,Mercado2026}.

\subsection{Halo identification}\label{sec:halo_finding}

We identify dark matter halos and characterize their properties with the AMIGA Halo Finder \citep[AHF;][]{Knollmann2009}.\footnote{\url{http://popia.ft.uam.es/AHF/}} AHF locates halos as spherical regions and sets their virial radius, $\rvir$, to the radius at which the mean enclosed density equals $\Delta(z)\,\rho_{\rm m}(z)$. The virial mass follows directly as the total mass enclosed within $\rvir$,

\begin{equation}
    \mvir = \frac{4}{3}\pi \Delta(z) \rho_{\rm m}(z) \rvir^3,
\end{equation}

where $\rho_{\rm m}(z)$ is the cosmic mean matter density at redshift $z$ and $\Delta(z)$ is the redshift-dependent virial overdensity defined by \citet{Bryan1998}.

AHF further organizes halos into groups, designating the most massive halo in each group as a ``central'' and any remaining group members as ``satellites.'' For each satellite, AHF truncates the halo at a radius identified from a local maximum in its radial density profile. We define a single halo radius and mass for every object as follows. For centrals, we adopt the virial quantities,

\begin{equation}
    r^{\rm centrals}_{\rm halo} \equiv \rvir, \quad M^{\rm centrals}_{\rm halo} \equiv \mvir,
\end{equation}

and, for satellites, the AHF truncation radius, $r_{\rm t}$, together with the total dark matter plus baryonic mass enclosed within it,

\begin{equation}
    r^{\rm satellites}_{\rm halo} \equiv r_{\rm t}, \quad M^{\rm satellites}_{\rm halo} \equiv M_{\rm tot}(<r_{\rm t}).
\end{equation}

\subsection{Galaxy sample selection and definitions}\label{sec:sample_selection}

We use the $z=0$ FIREbox galaxy catalog of \citet{Moreno2022}, which contains 1720 galaxies spanning stellar masses from $\mstar \approx 4.6\times10^6\,\msun$ to $\mstar \approx 5.5\times10^{11}\,\msun$ and a broad range of environments. Following \citet{Moreno2022}, every galaxy in this catalog contains at least 128 gravitationally bound star particles. Unless otherwise stated, all galaxy properties are measured using gravitationally bound particles associated with each halo.

Following \citet{Mowla2019}, we define $r_\star^{80}$ as the three-dimensional radius that contains 80 percent of the stellar mass bound to the dark matter halo, and we define the stellar mass of each galaxy as the bound stellar mass within this radius,

\begin{equation}
    \mstar \equiv \mstar(<r_{\star}^{80}).
\end{equation}
We measure $r_{\star}^{80}$ about each galaxy's stellar center, which we locate with a shrinking-sphere algorithm \citep{Power2003}.

AHF bound-particle lists are inclusive: the list for a host halo also contains the particles of its subhalos, so a host's ``bound'' stars include those of its satellites. In some cases, mostly at $\mstar\gtrsim10^{10}\,\msun$, this includes a companion of comparable stellar mass, which inflates the measured size and produces spuriously elongated shapes. We therefore assign each star particle to the smallest catalog galaxy whose bound list contains it, and remove from every galaxy the stars that belong to a smaller catalog galaxy, before computing $\mstar$, the stellar center, $r_{\star}^{80}$, shapes, and aperture-matched formation times.

To examine how environmental trends depend on galaxy mass, we divide the sample into three stellar-mass bins with fixed edges at $10^8$ and $10^{10}\,\msun$: a low-mass bin with $\mstar < 10^8\,\msun$ ($N=800$), an intermediate-mass bin with $10^8 \leq \mstar < 10^{10}\,\msun$ ($N=775$), and a high-mass bin with $\mstar \geq 10^{10}\,\msun$ ($N=145$). Comparing results across these bins lets us test how the connections between environment, formation time, and shape change with stellar mass. Our partial correlations also control for $\log\mstar$ within each bin (Section~\ref{sec:partial_correlation_application}), and Appendix~\ref{app:continuous_mass} shows that our conclusions do not depend on the choice of bin edges.

We hold these mass edges fixed throughout the paper, so a given bin always refers to the same stellar-mass range. The galaxies included in each bin, however, depend on the analysis. The SFH analyses (Section~\ref{sec:results_sfhs}) use all 1720 galaxies. The shape analyses require at least 300 star particles in the shape aperture (Section~\ref{sec:shape_measurements}), which removes all galaxies below $\mstar\approx10^{7.2}\,\msun$ and therefore restricts the low-mass bin to a narrower mass range. We report the sample size alongside each result.

\subsection{Calculating star formation histories of simulated galaxies}\label{sec:sfhs}

We construct each galaxy's star formation history from the formation times of its surviving star particles at $z=0$, using the star selection described in Section~\ref{sec:sample_selection}. The cumulative SFH is the fraction of this present-day stellar mass that had formed by each lookback time. We weight each star particle by its present-day mass and do not correct for stellar mass loss. Because it includes these stars regardless of whether they formed in the main progenitor or in an accreted system, it records when the present-day stars formed rather than when they were assembled into the galaxy. These are therefore archaeological SFHs, analogous to those inferred from the resolved stellar populations of Local Group galaxies. Observations of more massive or more distant galaxies rarely constrain the full lifetime SFH in this way, but we use archaeological SFHs for all simulated galaxies for consistency.

From these cumulative star formation histories, we define three formation-time metrics: $\tau_{30}$, $\tau_{50}$, and $\tau_{90}$. These quantities correspond to the lookback times by which a galaxy had formed 30, 50, and 90 percent of its present-day stellar mass, respectively. Larger values of $\tau$ therefore indicate earlier star formation. $\tau_{30}$ traces the earliest phase of stellar mass assembly, $\tau_{50}$ traces the median formation time of the stellar population, and $\tau_{90}$ is especially sensitive to recent star formation and late-time stellar mass growth.

We also examine radially resolved star formation histories by selecting stars according to their present-day distance from the stellar center. In this case, we divide each galaxy into three spherical shells using the three-dimensional bound stellar half-mass radius, $\rhalf$, as the radial scale: $0<r/\rhalf<1$, $1<r/\rhalf<2$, and $2<r/\rhalf<4$. We measure $\rhalf$ and the stellar radii in the same way as $r_\star^{80}$: about the stellar center, after removing substructure stars (Section~\ref{sec:sample_selection}). We then recompute the cumulative star formation history and formation-time metrics in each radial region. These radial measurements allow us to test whether environmental trends in formation time are concentrated in the inner galaxy or distributed throughout the stellar body.

\subsection{Measuring shapes of simulated galaxies}\label{sec:shape_measurements}

To quantify galaxy structure, we measure the three-dimensional stellar shape of each galaxy using an iterative shape-tensor approach \citep[e.g.,][]{BailinSteinmetz2005,Allgood2006} with a reduced inertia tensor \citep{Chua2019}. This method has previously been applied to FIRE galaxies by \citet{Xia2026} and \citet{Klein2026}. Inertia-tensor axis ratios can be sensitive to outliers, such as substructure missed by the removal step, for which robust estimators such as the minimum covariance determinant provide an alternative \citep{Geng2026}; we limit this sensitivity by removing the stars of resolved substructure (Section~\ref{sec:sample_selection}).

Shape measurements use the gravitationally bound star particles within a sphere of radius $r_\star^{80}$ centered on the stellar center of each galaxy (Section~\ref{sec:sample_selection}). We tie both the center and the size of the aperture to the stars rather than to the dark matter halo.

We define the reduced inertia tensor as

\begin{equation}
    S_{ij} =
    \frac{\sum_n m_n r_{{\rm ell},n}^{-2} x_{n,i} x_{n,j}}
    {\sum_n m_n},
    \label{eq:shape_tensor}
\end{equation}

where the sum is over star particles in the current aperture, $m_n$ is the stellar particle mass, and $x_{n,i}$ and $x_{n,j}$ are the particle coordinates along the $i$ and $j$ spatial directions. The factor $r_{{\rm ell},n}^{-2}$ downweights particles at large ellipsoidal radii, reducing the sensitivity of the shape measurement to the sparse outer regions of the galaxy. The ellipsoidal radius is defined as

\begin{equation}
    r_{\rm ell} =
    \sqrt{
    x^2
    + \frac{y^2}{(b/a)^2}
    + \frac{z^2}{(c/a)^2}
    },
    \label{eq:ellipsoidal_radius}
\end{equation}

where $a\geq b\geq c$ are the major, intermediate, and minor axes of the current best-fit ellipsoid.

The shape measurement proceeds iteratively. We first re-center the aperture particles on their weighted center of mass and compute the reduced inertia tensor for all of them. Diagonalizing this tensor gives an initial estimate of the principal axes and their orientations. The particle distribution is then rotated into the principal-axis frame, the ellipsoid is rescaled to preserve its volume, and particles outside the current ellipsoidal aperture are removed. We repeat this procedure until the axis ratios $b/a$ and $c/a$ change by less than $10^{-3}$ between iterations, or until a maximum of 30 iterations is reached. The axis ratios we report are those of the converged ellipsoid. To assess how well each measurement is resolved, we then count the star particles within the ellipsoidal radius enclosing the inner 90 percent of the remaining stellar mass, which excludes the small number of outlying particles near the edge of the converged aperture.

From the converged ellipsoid, we retain the intermediate-to-major axis ratio, $b/a$, the minor-to-major axis ratio, $c/a$, and the triaxiality parameter,

\begin{equation}
    T = \frac{a^2-b^2}{a^2-c^2}.
    \label{eq:triaxiality}
\end{equation}

The axis ratio $c/a$ measures the thickness or flattening of the stellar distribution along the minor axis, while $b/a$ measures the degree of in-plane elongation. The triaxiality parameter distinguishes oblate-like systems, with $T\rightarrow0$, from prolate-like systems, with $T\rightarrow1$.

We measure shapes for several stellar-population and weighting choices. In this work, our fiducial shape measurements use all bound star particles within the $r_\star^{80}$ aperture and weight particles by stellar mass. We also consider luminosity-weighted shapes, in which the particle masses in Equation~\ref{eq:shape_tensor} are replaced by SDSS $r$-band luminosities estimated from stellar population synthesis lookup tables using each particle's age, metallicity, and mass, assuming a \citet{Chabrier2003} initial mass function. Although FIREbox assumes a \citet{Kroupa2002} initial mass function, the two differ mainly in the abundance of low-mass stars, which contribute little $r$-band light, and any constant offset in the mass-to-light ratio cancels in Equation~\ref{eq:shape_tensor}, which normalizes by the total weight. These luminosity-weighted measurements provide a closer analogue to shapes inferred from photometric observations, while the mass-weighted measurements trace the intrinsic stellar mass distribution \citep{Klein2026}. %

We retain a shape measurement only if at least 300 star particles lie within the inner-90-percent ellipsoid described above. We apply this cut to limit the effect of small-number particle noise on the inferred axis ratios. We adopt this relatively strict threshold because many galaxies in our low-mass bin lie relatively close to the FIREbox resolution limit, with only a few hundred star particles each. \citet{Geng2026} find that inertia-tensor axis ratios become essentially unbiased once the number of particles reaches $\sim10^3$, so our threshold balances measurement precision against sample size in the low-mass bin.

Because star particles have nearly equal masses, the number of particles in the aperture tracks stellar mass closely, so this cut acts almost exactly as a stellar-mass threshold: for the mass-weighted shapes, every galaxy with $\mstar\geq10^{7.18}\,\msun$ passes it and every galaxy below fails. The intermediate- and high-mass bins are therefore complete, while 410 of the 800 low-mass galaxies retain a reliable shape (402 for the luminosity-weighted shapes, for which the cut is slightly less sharp). The shape results for the low-mass bin thus describe galaxies with $10^{7.2}\lesssim\mstar<10^8\,\msun$ (median $\log_{10}(\mstar/\msun)=7.58$, compared with 7.19 for the full bin), whereas the SFH results for this bin include all 800 galaxies.

We also assign each galaxy to a coarse morphological class based on its location in the $b/a$--$c/a$ plane, following \citet{vanderWel2014b}. Galaxies are classified as disky, spheroidal, or elongated according to their relative axis ratios. Disky systems occupy the region near flattened oblate configurations, spheroidal systems are close to round in all three axes, and elongated systems correspond to more prolate or triaxial configurations. These classifications provide a useful summary of the shape distribution, although Section~\ref{sec:results_shape_classes} shows that discrete morphology classes do not always preserve the environmental trends present in the continuous axis ratios.

\begin{figure}
    \centering
    \includegraphics[width=1.0\linewidth, trim = 0 0 0 0]{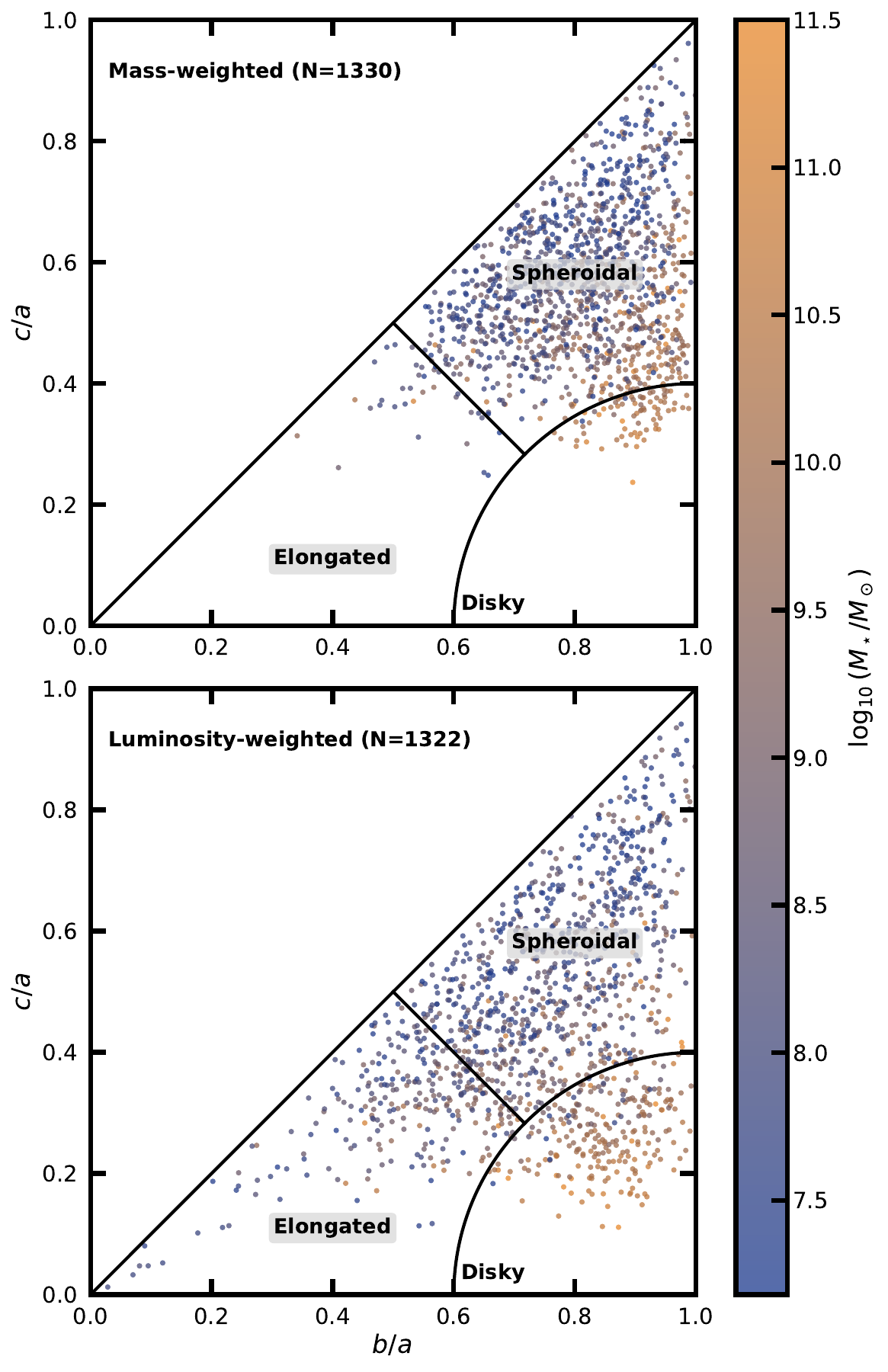}
    \caption{
    The Disky/Spheroidal/Elongated classification boundaries (black lines; Section~\ref{sec:shape_measurements}) in the $b/a$--$c/a$ plane, with the resolved FIREbox galaxies color-coded by $\log_{10}(M_\star)$. Only galaxies with reliable shape measurements (at least 300 star particles; Section~\ref{sec:shape_measurements}) are shown, so $N$ is smaller than the full sample of 1720 and differs slightly between the two weightings. \textbf{Top:} fiducial mass-weighted shapes. \textbf{Bottom:} luminosity-weighted shapes, used as a robustness check throughout this work.
    }
    \label{fig:shape_classification_diagram}
\end{figure}

Figure~\ref{fig:shape_classification_diagram} shows these classification boundaries directly in the $b/a$--$c/a$ plane, together with the resolved FIREbox population, color-coded by stellar mass, for both the mass-weighted (top) and luminosity-weighted (bottom) shape measurements. Most galaxies fall in the Spheroidal region in both weighting schemes, but the luminosity-weighted population extends further into the Elongated and Disky regions, particularly at higher stellar mass. This reflects the same physical distinction discussed above: luminosity weighting emphasizes younger, more rotationally supported stellar populations, which more readily populate flattened configurations, while mass weighting traces the older, dynamically hotter stellar component that dominates the total stellar mass.

We note that seven luminosity-weighted shapes lie near $b/a\approx c/a\approx0$ in the lower-left corner of the bottom panel. In these galaxies, the $r$-band light within the aperture is dominated by a small number of very young star particles, so the luminosity-weighted fit traces those particles rather than the overall stellar distribution; their mass-weighted shapes are not unusual. We retain them in the sample, but excluding them changes the luminosity-weighted correlations reported in Section~\ref{sec:results} by less than 0.01.

\subsection{Quantifying environment}\label{sec:environment}

We quantify each galaxy's local environment using a simulation-adapted Perturbation Index (PI), inspired by the tidal-strength formalism of \citet{Verley2007}. This metric estimates the cumulative strength of external tidal perturbations relative to the internal binding force of the galaxy being perturbed. We define

\begin{equation}
    {\rm PI} \equiv \sum_i 
    \left(\frac{M_i}{\mhalo}\right)
    \left(\frac{D_{\rm halo}}{d_i}\right)^3
    \propto
    \sum_i \frac{F_{{\rm tidal},i}}{F_{\rm bind}},
    \label{eq:PI}
\end{equation}

where $M_i$ is the halo mass of the $i$th neighboring system, $d_i$ is the three-dimensional distance between the galaxy and that neighbor, $\mhalo$ is the halo mass of the galaxy whose environment is being measured, and $D_{\rm halo}=2\rhalo$ is its halo diameter. We compute this sum over all other galaxies in our catalog (Section~\ref{sec:sample_selection}), so it includes the host of a satellite and the resolved satellites of a central. Because FIREbox is a periodic volume, we use the nearest periodic image of each neighbor, so that each neighbor is counted once, and we include neighbors within a three-dimensional separation of 10\,Mpc, the scale to which PI is most sensitive \citep{Mercado2025}. Extending the sum to the full box changes PI by a median of 0.003\,dex.

This definition follows the scaling used by \citet{Verley2007}, while replacing projected separations and optical diameters with the three-dimensional halo separations, halo masses, and halo diameters available in the simulation. Larger values of PI indicate stronger present-day tidal perturbations from neighboring systems.

Within this normalization, ${\rm PI}=1$ provides a convenient reference value, roughly where the estimated cumulative tidal influence of neighboring systems becomes comparable to the galaxy's internal binding force. No galaxy has a PI of exactly 1, and we use this value to define the high-PI (${\rm PI}>1$) and low-PI (${\rm PI}<1$) populations compared throughout Sections~\ref{sec:results_sfhs},~\ref{sec:results_shapes}, and~\ref{sec:results_shape_classes}.

A detailed discussion of the PI calculation, its physical interpretation, and its connection to other environment metrics in FIREbox is presented in \citet{Mercado2025}. In that work, we used a similar radius-based normalization of PI ($\rhalo$ in place of $D_{\rm halo}$ in Equation~\ref{eq:PI}) and showed that the metric is most sensitive to neighboring structure on scales of roughly 1--10 Mpc. The diameter-based convention adopted here shifts the normalization by a constant factor of $2^3$, or 0.90 dex in $\log_{10}({\rm PI})$, and therefore does not change the relative environmental ranking of galaxies or the trends reported in \citet{Mercado2025}. We adopt the diameter-based definition in this work because it more closely follows the original \citet{Verley2007} convention and provides a convenient reference scale for comparing high- and low-PI populations.

\subsection{Partial-correlation analysis}
\label{sec:partial_correlation}

To test whether the connection between environment and galaxy shape is linked to formation time, we use a partial-correlation analysis. This choice is motivated by the tidal-stirring picture, in which repeated tidal encounters can both truncate star formation and dynamically heat stellar orbits, transforming rotating dwarf irregulars into pressure-supported spheroidals \citep{Mayer2001,Kazantzidis2011}. In this framework, formation time and shape are expected to be linked, so we test whether formation time statistically accounts for the environment--shape relation. We first describe the method for three generic quantities, then how we interpret its results, and finally how we apply it to FIREbox galaxies.

\subsubsection{Partial correlations}\label{sec:partial_correlation_method}

Consider three quantities $X$, $Y$, and $Z$. The Spearman rank correlation, $\rho(X,Y)$, measures the strength of the monotonic correlation between $X$ and $Y$. The partial correlation, $\rho(X,Y \mid Z)$, measures the correlation between $X$ and $Y$ that remains when $Z$ is held fixed. We compute it by regressing the ranks of $X$ and the ranks of $Y$ separately against the ranks of $Z$ using least-squares regression, and then computing the Pearson correlation coefficient between the two sets of residuals \citep[e.g.,][]{Baba2004}. Because the calculation uses ranks, it does not assume linear relations among the quantities. Additional control variables are included by regressing against the ranks of all of them simultaneously. Partial Spearman correlations have previously been used to disentangle the interdependence of galaxy morphology, star formation, and environment \citep{Bait2017} and to compare the dependence of quenching on global and spatially resolved galaxy properties \citep{Bluck2020}. In our analysis, every correlation controls for stellar mass (Section~\ref{sec:partial_correlation_application}). The method therefore consists of comparing the mass-controlled correlation, $\rho(X,Y \mid \log\mstar)$, with the partial correlation at fixed $Z$ and stellar mass, $\rho(X,Y \mid Z,\log\mstar)$, so that the difference isolates how much of the correlation between $X$ and $Y$ is accounted for by their correlations with $Z$. Because $Z$ can account for the $X$--$Y$ correlation only if it is correlated with both $X$ and $Y$, we also report $\rho(X,Z \mid \log\mstar)$ and $\rho(Y,Z \mid \log\mstar)$. We estimate uncertainties by bootstrap resampling the galaxies in each stellar-mass bin 2000 times and report the 16th--84th and 2.5th--97.5th percentile ranges of the resulting correlations.

\subsubsection{Interpreting the partial correlations}\label{sec:partial_correlation_interpretation}

\begin{figure*}[t]
    \centering
    \includegraphics[width=0.95\textwidth, trim = 0 0 0 0]{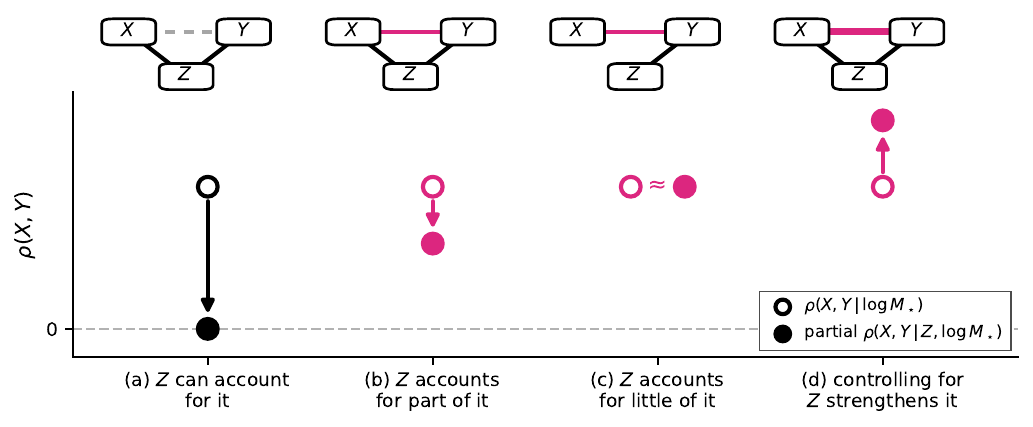}
    \caption{
    Schematic of the partial-correlation test for three quantities $X$, $Y$, and $Z$. Open circles show the Spearman correlation at fixed stellar mass, $\rho(X,Y \mid \log\mstar)$, filled circles show the partial correlation after also controlling for $Z$, and arrows point from the former to the latter. Black (pink) indicates that the $X$--$Y$ correlation vanishes (remains) at fixed $Z$. Panels (a)--(d) show the four outcomes described in Section~\ref{sec:partial_correlation_interpretation}. \textbf{Diagrams above each case:} the corresponding correlations; the $X$--$Y$ link is dashed when it vanishes at fixed $Z$, and drawn heavier in (d). Correlation values are illustrative.
    }
    \label{fig:partial_correlation_schematic}
\end{figure*}

Figure~\ref{fig:partial_correlation_schematic} illustrates how we interpret the comparison between $\rho(X,Y \mid \log\mstar)$ and $\rho(X,Y \mid Z,\log\mstar)$. If the partial correlation drops to a value consistent with zero (Figure~\ref{fig:partial_correlation_schematic}a), the correlation between $X$ and $Y$ can be statistically accounted for by their shared correlation with $Z$. If it decreases but remains significantly different from zero (Figure~\ref{fig:partial_correlation_schematic}b), $Z$ accounts for only part of the correlation, and a residual correlation between $X$ and $Y$ remains at fixed $Z$. The size of the decrease therefore indicates how much of the correlation $Z$ accounts for, while whether the partial correlation is consistent with zero, judged from its $p$-value and bootstrap range, indicates whether a correlation remains at fixed $Z$. Two further outcomes are possible. If the partial correlation is close to the mass-controlled correlation (Figure~\ref{fig:partial_correlation_schematic}c), $Z$ accounts for little of the correlation, as expected when $Z$ is uncorrelated with $X$ or with $Y$. If the partial correlation is larger than the mass-controlled correlation (Figure~\ref{fig:partial_correlation_schematic}d), controlling for $Z$ strengthens the correlation between $X$ and $Y$, which can occur when $Z$ is correlated with one of the two quantities in a way that partly masks their relation.

\subsubsection{Application to FIREbox galaxies}\label{sec:partial_correlation_application}

We treat PI as the environmental variable, $\tau_{90}$ as the formation-time variable, and $c/a$ as the shape variable. For each stellar-mass bin, we compute the Spearman correlations among these three quantities at fixed stellar mass, $\rho({\rm PI},\tau_{90} \mid \log\mstar)$, $\rho({\rm PI},c/a \mid \log\mstar)$, and $\rho(\tau_{90},c/a \mid \log\mstar)$, which quantify, respectively, the correlation between present-day environment and formation time, between present-day environment and shape, and between formation time and shape. We then apply the test in two directions. For $\rho({\rm PI},c/a \mid \tau_{90},\log\mstar)$, with $X={\rm PI}$, $Y=c/a$, and $Z=\tau_{90}$, a partial correlation consistent with zero would indicate that the PI--shape relation can be accounted for by formation time. For $\rho(\tau_{90},c/a \mid {\rm PI},\log\mstar)$, with $X=\tau_{90}$, $Y=c/a$, and $Z={\rm PI}$, a partial correlation that remains close to the mass-controlled value would indicate that the formation time--shape relation is not a by-product of environment. Because the partial correlation is symmetric in $X$ and $Y$, the two tests differ only in which variable is held fixed.

Controlling for $\log\mstar$ is necessary because the stellar-mass bins are broad, and the high-mass bin is unbounded above. As a result, PI and stellar mass can remain correlated within a bin. For the galaxies with reliable mass-weighted shapes, the within-bin Spearman correlation between PI and $\log\mstar$ is small ($\rho({\rm PI},\log\mstar)=-0.04$, $0.04$, and $0.09$ in the low-, intermediate-, and high-mass bins), but because both $\tau_{90}$ and $c/a$ depend strongly on stellar mass, we control for it explicitly. We also repeat the analysis with halo mass as an additional control variable (Section~\ref{sec:results_halo_mass}) and for central galaxies alone (Section~\ref{sec:results_shape_sfh}).

\textit{We emphasize that this is a correlational test and does not establish a causal pathway.} In particular, it cannot distinguish whether PI traces direct tidal interactions that influence both star formation and shape, or whether PI and formation time share a common origin in the assembly history of the surrounding large-scale structure.

\begin{figure*}
    \centering
    \includegraphics[width=1.0\textwidth, trim = 0 0 0 0]{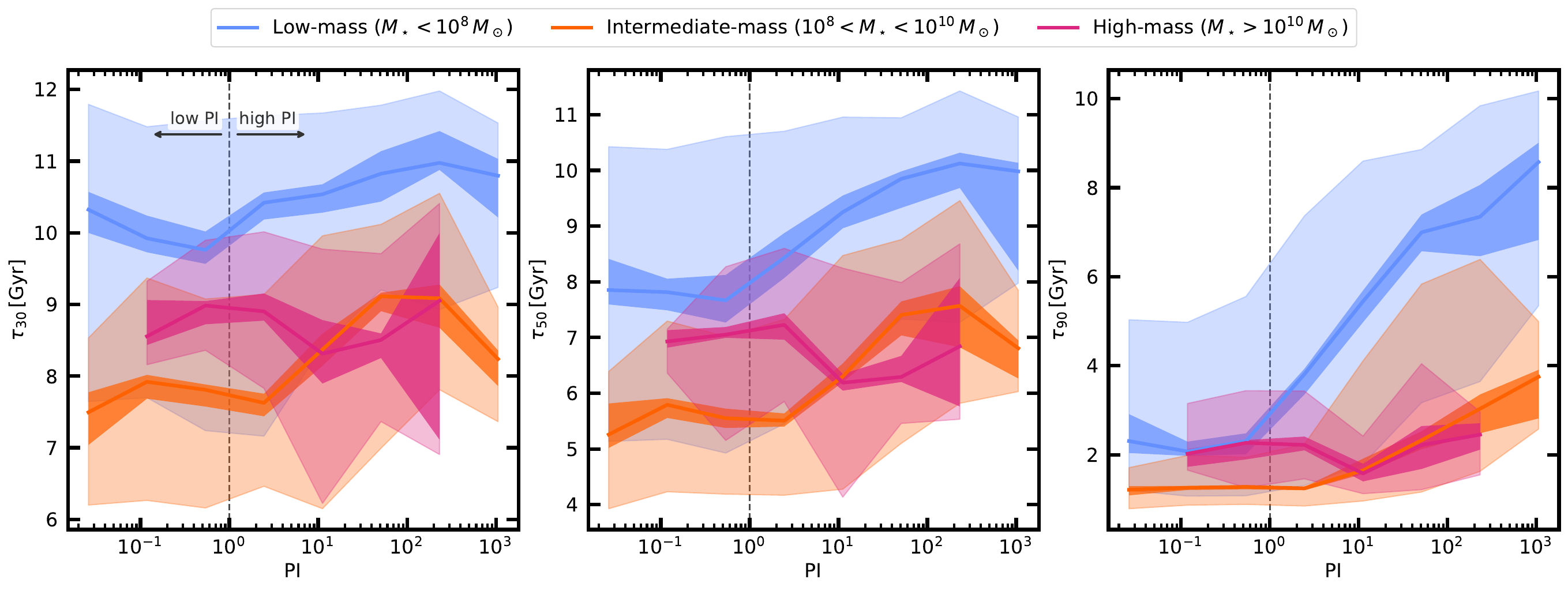}
    \caption{
    Formation-time metrics as a function of Perturbation Index for the low- (blue), intermediate- (orange), and high-mass (magenta) bins. Lines show median trends in bins of PI, dark shaded bands show bootstrap 16th--84th percentile ranges of the median, and light shaded regions show the 16th--84th percentile ranges of the galaxy distribution. \textbf{Left:} $\tau_{30}$; \textbf{middle:} $\tau_{50}$; \textbf{right:} $\tau_{90}$, the lookback times by which galaxies formed 30, 50, and 90 percent of their present-day stellar mass. The vertical dashed line marks ${\rm PI}=1$, which separates the low- and high-PI regimes (Section~\ref{sec:environment}). Low-mass galaxies have earlier $\tau_{30}$ and $\tau_{50}$ values overall. In the low- and intermediate-mass bins, formation times are earlier at higher PI, most clearly in $\tau_{90}$ for the low-mass bin; the high-mass bin shows no clear trend.
    }
    \label{fig:tau_vs_pi}
\end{figure*}

\begin{figure*}
    \centering
    \includegraphics[width=0.75\textwidth, trim = 0 0 0 0]{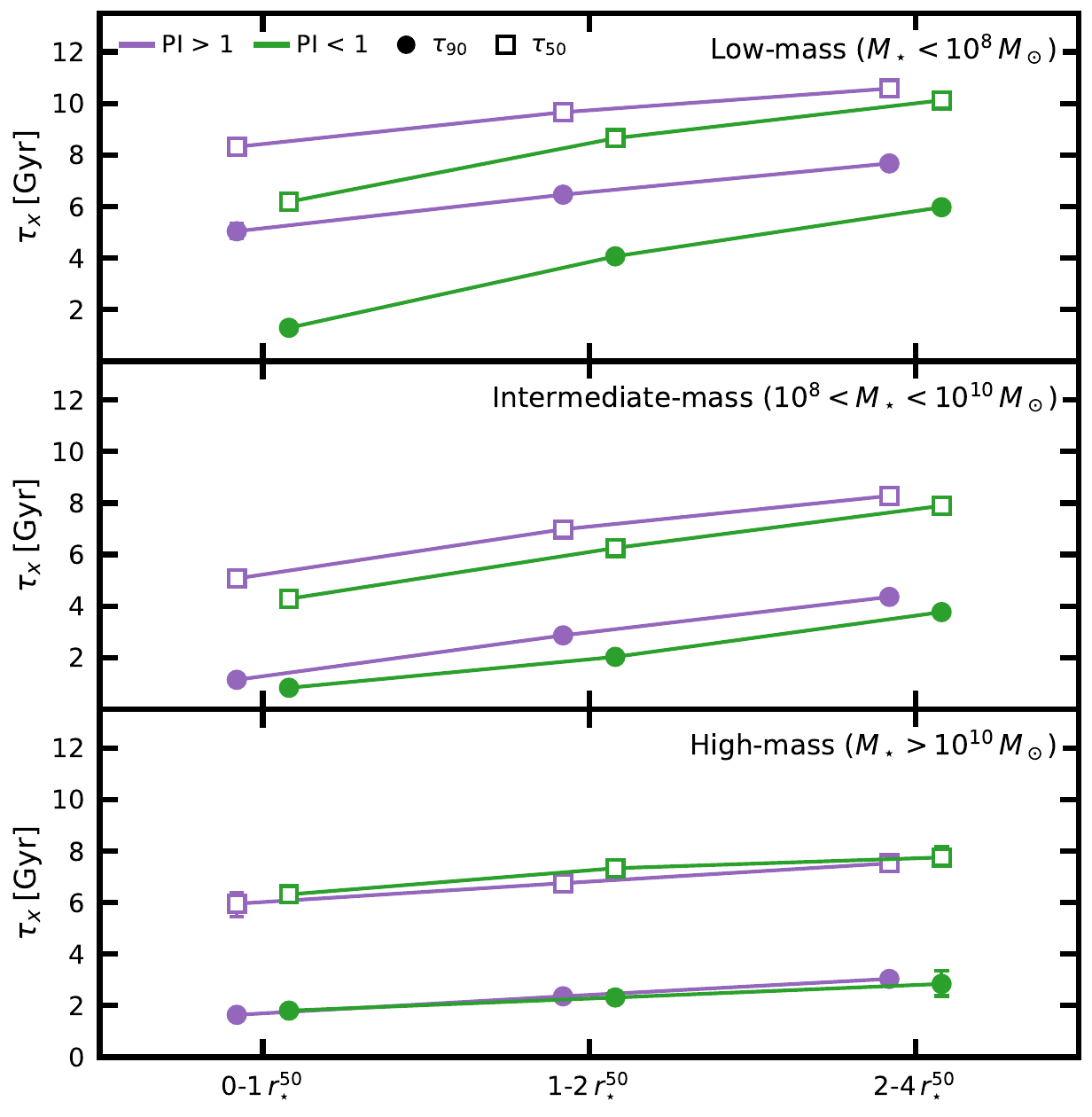}
    \caption{
    Median formation times of galaxies with ${\rm PI}>1$ (purple) and ${\rm PI}<1$ (green) in the low- (top), intermediate- (middle), and high-mass (bottom) bins. Filled circles show $\tau_{90}$ and open squares show $\tau_{50}$. Points show stars currently located at $0-1\rhalf$, $1-2\rhalf$, and $2-4\rhalf$, after removing substructure stars. Error bars show the 16th--84th percentile range of the median from bootstrap resampling of the galaxies in each subsample; Figure~\ref{fig:tau_vs_pi} shows the galaxy-to-galaxy scatter of the whole-galaxy values. A galaxy contributes to a shell only if it has at least 10 star particles there, which removes at most 12 of the 800 low-mass galaxies and none in the higher-mass bins. In the low- and intermediate-mass bins, high-PI galaxies have larger $\tau_{50}$ and $\tau_{90}$ than low-PI galaxies in every shell, with the largest difference in the low-mass bin for stars within $\rhalf$. In the high-mass bin, $\tau_{90}$ is similar for the two populations in every shell.
    }
    \label{fig:sfh_pi_split}
\end{figure*}

\section{Results}\label{sec:results}

\subsection{Environmental trends in formation times}\label{sec:results_sfhs}

We first examine whether a galaxy's present-day tidal environment is connected to when it formed its stars. As described in Section~\ref{sec:sfhs}, these formation times record when the present-day stars formed, not when they were assembled into the galaxy. Figure~\ref{fig:tau_vs_pi} shows $\tau_{30}$, $\tau_{50}$, and $\tau_{90}$ as a function of PI for the three stellar-mass bins. The medians are computed in logarithmic bins of PI that contain at least 5 galaxies; most bins contain tens to $\sim180$ galaxies, and the smallest are the highest-PI bins of the intermediate- and high-mass samples, with 5 and 8 galaxies, respectively.

The strongest overall dependence in Figure~\ref{fig:tau_vs_pi} is with stellar mass. The low-mass bin has the largest values of $\tau_{30}$ and $\tau_{50}$ across most of the PI range: these galaxies formed a larger fraction of their present-day stellar mass at early times than galaxies in the intermediate- and high-mass bins. The mass ordering is not monotonic: at ${\rm PI}\lesssim10$, the intermediate-mass bin has the smallest values of all three formation-time metrics, so these galaxies formed their stars latest.

In the low- and intermediate-mass bins, we also find a trend with environment: galaxies in more strongly perturbing environments have larger values of $\tau_{30}$, $\tau_{50}$, and $\tau_{90}$. The trend strengthens from $\tau_{30}$ to $\tau_{90}$, with Spearman correlations $\rho({\rm PI},\tau)=0.15$, $0.24$, and $0.44$ in the low-mass bin and $0.22$, $0.27$, and $0.38$ in the intermediate-mass bin (all $p<10^{-4}$). The high-mass bin shows no significant trend in any of the three metrics ($|\rho|\leq0.10$, $p\geq0.24$). The clearest environmental signal appears in $\tau_{90}$ for the low-mass bin. Low-mass galaxies at low PI have median $\tau_{90}$ values of only a few Gyr, while those at higher PI have substantially larger $\tau_{90}$ values, indicating that they completed most of their stellar mass growth earlier. The strong dependence of $\tau_{90}$ on PI therefore suggests that the present-day environment leaves its most pronounced imprint on the later-time star formation of low-mass galaxies.

The intermediate-mass bin shows a weaker version of the same trend: its median formation times increase with PI mainly at ${\rm PI}\gtrsim10$, and the change across the PI range is smaller than in the low-mass bin, especially for $\tau_{90}$. In both higher-mass bins, the 16th--84th percentile regions overlap substantially, indicating significant object-to-object scatter within each mass bin at fixed PI.

To see where within galaxies the environmental difference appears, we divide each stellar-mass bin at ${\rm PI}=1$ and compare the formation times of the two populations in three spherical shells defined by the present-day stellar half-mass radius (Section~\ref{sec:sfhs}). Figure~\ref{fig:sfh_pi_split} shows the median $\tau_{50}$ and $\tau_{90}$ in each shell. In every stellar-mass bin and for both populations, $\tau_{50}$ and $\tau_{90}$ increase with radius: stars currently located at larger radii formed earlier on average.

Within the low- and intermediate-mass bins, high-PI galaxies formed their stars earlier than low-PI galaxies in every shell. In the low-mass bin, this difference is largest in the inner regions: the $\tau_{90}$ difference between high- and low-PI galaxies within $\rhalf$ is about twice that in the $2-4\rhalf$ shell, and $\tau_{50}$ follows the same pattern. In the intermediate-mass bin, the difference is smaller and spread more evenly across the shells. In the high-mass bin, the two populations have similar formation times in every shell: the median $\tau_{90}$ differs by less than $0.2$\,Gyr and the median $\tau_{50}$ by at most $\approx0.6$\,Gyr, much smaller than the differences in the low-mass bin. This is consistent with the absence of a significant correlation between formation time and PI in this bin (Figure~\ref{fig:tau_vs_pi}), although the low-PI subsample here contains only 36 galaxies.

Because the shells are defined using present-day stellar positions, they do not trace where stars formed: radial redistribution can move stars away from their birth radii. The radial trends therefore show that stars currently in the inner regions of low-mass, high-PI galaxies are older on average than those of low-PI galaxies of similar mass.

Together, Figures~\ref{fig:tau_vs_pi} and~\ref{fig:sfh_pi_split} show that stellar mass sets the broad sequence in formation time, while present-day tidal environment is correlated with additional variation below $10^{10}\,\msun$. This variation is strongest for the late-time star formation of low-mass galaxies and, in that bin, for the stars currently in the inner regions.

\subsection{Environmental trends in continuous galaxy shapes}
\label{sec:results_shapes}

\begin{figure*}
    \centering
    \includegraphics[width=1.0\textwidth, trim = 0 0 0 0]{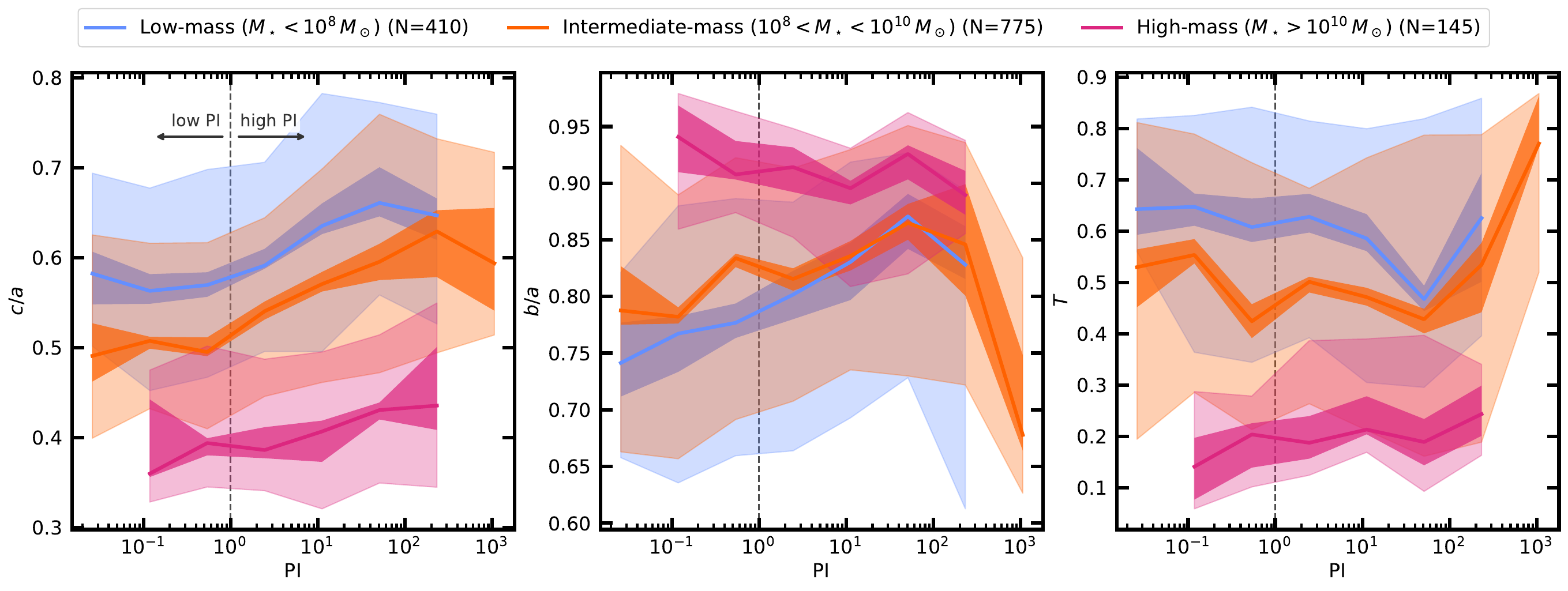}
    \caption{
    Mass-weighted three-dimensional stellar shape parameters as a function of Perturbation Index for the low- (blue), intermediate- (orange), and high-mass (magenta) bins, measured from the bound star particles within $r_\star^{80}$ of the stellar center. Lines show running medians, dark shaded bands show bootstrap 16th--84th percentile ranges of the median, and light shaded regions show the 16th--84th percentile ranges of the galaxy distribution. \textbf{Left:} minor-to-major axis ratio, $c/a$; \textbf{middle:} intermediate-to-major axis ratio, $b/a$; \textbf{right:} triaxiality parameter, $T$. The vertical dashed line marks ${\rm PI}=1$. In all three bins, $c/a$ increases with PI; in the low- and intermediate-mass bins, $b/a$ also increases, more weakly. $T$ depends primarily on stellar mass and only weakly on PI.
    }
    \label{fig:shape_vs_pi}
\end{figure*}

We next ask whether the environmental trends in formation time are accompanied by corresponding changes in galaxy structure. Figure~\ref{fig:shape_vs_pi} shows the median three-dimensional stellar shape parameters as a function of PI for the three stellar-mass bins, using the fiducial mass-weighted shapes described in Section~\ref{sec:shape_measurements}. The three panels show the minor-to-major axis ratio, $c/a$, the intermediate-to-major axis ratio, $b/a$, and the triaxiality parameter, $T$ (Equation~\ref{eq:triaxiality}).

The shape parameters depend strongly on stellar mass: the median $c/a$ decreases from $0.60$ in the low-mass bin to $0.54$ and $0.41$ in the intermediate- and high-mass bins, so more massive galaxies are thinner along their minor axes. This trend toward flatter shapes at higher mass agrees with previous measurements of FIREbox galaxy shapes \citep{Klein2026} and is qualitatively consistent with observations, in which dwarf galaxies are thick and slightly triaxial \citep{KadoFong2020} and become thicker toward lower luminosity \citep{Carlsten2021}. Within each bin, the clearest environmental dependence appears in $c/a$. In the low-mass bin, the median $c/a$ increases from $\sim0.57$ at low PI to $\sim0.65$ at ${\rm PI}\gtrsim10$; in the intermediate-mass bin, it increases from $\sim0.50$ to $\sim0.60$; and in the high-mass bin, from $\sim0.36$ to $\sim0.44$. Within these bins, PI is nearly uncorrelated with $\log\mstar$ (Section~\ref{sec:partial_correlation_application}), so high- and low-PI galaxies in a given bin span similar stellar masses. The corresponding Spearman correlations are $\rho({\rm PI},c/a)=0.26$, $0.30$, and $0.19$ in the low-, intermediate-, and high-mass bins (Table~\ref{tab:partial_correlation_summary} gives the corresponding values at fixed stellar mass). The intermediate-to-major axis ratio, $b/a$, also increases with PI in both the low- and intermediate-mass bins ($\rho=0.22$ and $0.18$, respectively), from a median of $\approx0.76$ and $0.79$, respectively, at ${\rm PI}<0.1$ to $\approx0.85$ in both bins at ${\rm PI}\gtrsim10$, above which it levels off; the slight decrease in the highest-PI low-mass galaxies (${\rm PI}>100$) is based on only 15 galaxies, and the sharp change in the highest-PI intermediate-mass bin (${\rm PI}\sim10^3$) is based on only 5 galaxies. Because both $c/a$ and $b/a$ increase, low- and intermediate-mass galaxies in more strongly perturbing environments are more spherical, rather than only thicker. In the high-mass bin, $b/a$ remains $\approx0.9$ and shows no significant trend with PI, so the increase in $c/a$ there corresponds to oblate stellar distributions that are thicker along the minor axis. Because $c/a$ describes the overall stellar distribution within $r_\star^{80}$, a more prominent round central component could also contribute to this increase. The triaxiality parameter depends only weakly on PI ($|\rho({\rm PI},T)|\leq0.19$) and is instead primarily mass dependent, decreasing from $T\approx0.6$ in the low-mass bin to $T\approx0.2$ (oblate) in the high-mass bin.

The luminosity-weighted shapes show the same pattern, with $\rho({\rm PI},c/a)=0.32$, $0.29$, and $0.20$. The correlation is highly significant ($p<10^{-7}$) in the two lower-mass bins and marginally significant in the high-mass bin ($p=0.02$ for both weightings), which contains only 145 galaxies.

Overall, present-day tidal environment is correlated with galaxy shape within all three stellar-mass bins, primarily through $c/a$: galaxies in more strongly perturbing environments are more spherical at low and intermediate mass and thicker along the minor axis at high mass. In Section~\ref{sec:results_shape_classes}, we test whether this continuous trend is also recovered when the same galaxies are grouped into broad morphology classes.

\subsection{Environmental trends in morphology classifications}
\label{sec:results_shape_classes}

We apply the Disky/Spheroidal/Elongated classification described in Section~\ref{sec:shape_measurements} and compare the class fractions of low-PI (${\rm PI}<1$) and high-PI (${\rm PI}>1$) galaxies, using the same PI threshold as in Section~\ref{sec:results_sfhs}.

\begin{figure*}
    \centering
    \includegraphics[width=0.95\textwidth, trim = 0 0 0 0]{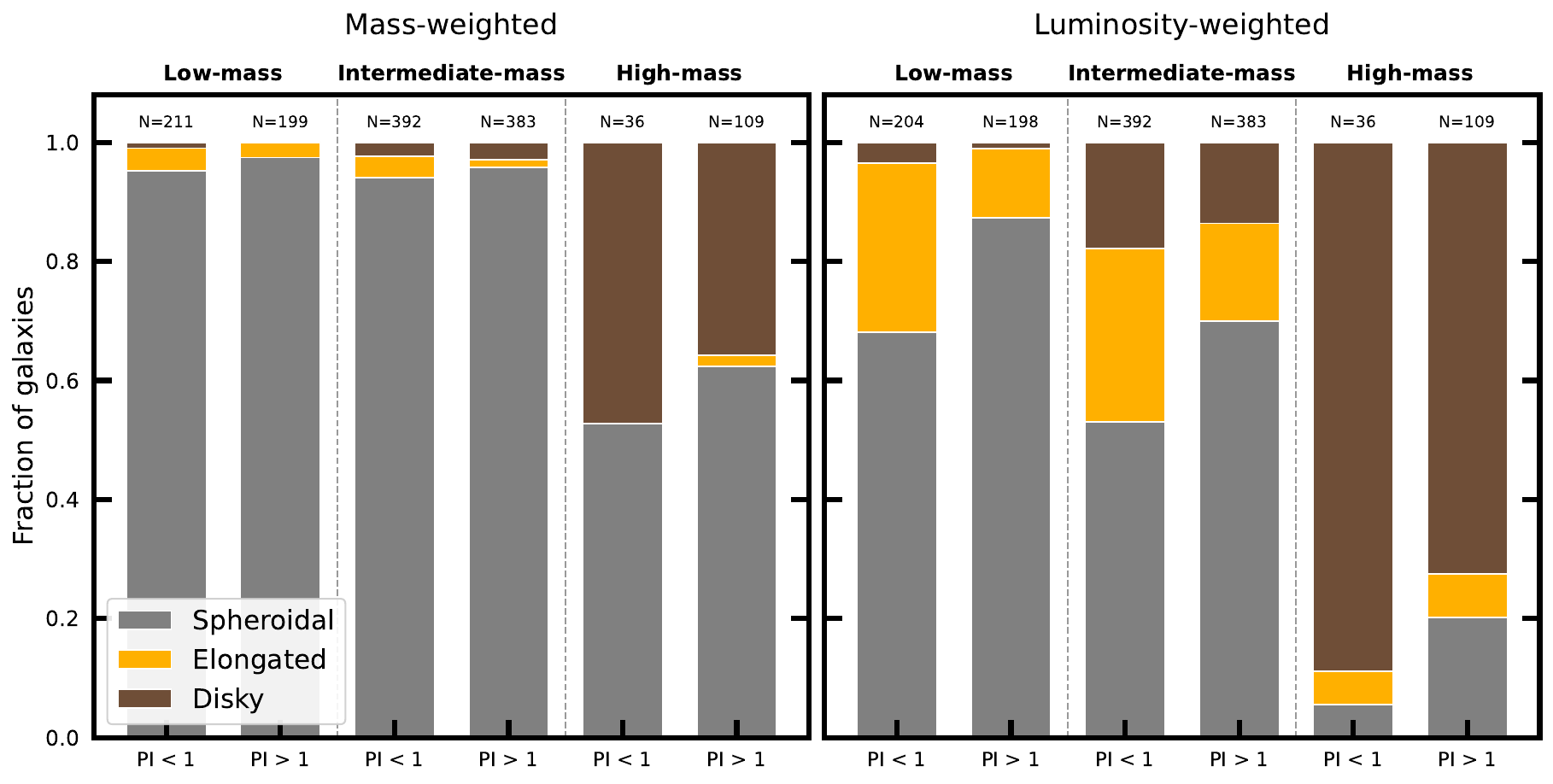}
    \caption{
    Morphological class fractions for low-PI (${\rm PI}<1$) and high-PI (${\rm PI}>1$) galaxies in each stellar-mass bin, based on each galaxy's location in the $b/a$--$c/a$ plane (Section~\ref{sec:shape_measurements}): Spheroidal (gray), Elongated (gold), and Disky (brown). Numbers above the bars give the number of galaxies in each group. \textbf{Left:} mass-weighted shapes; the fractions do not depend significantly on PI in any bin. \textbf{Right:} luminosity-weighted shapes; in the low- and intermediate-mass bins, high-PI galaxies are less likely to be classified as Elongated and more likely to be classified as Spheroidal, consistent with the continuous increase in $c/a$ with PI, while the high-mass bin is dominated by Disky systems and shows no significant PI dependence.
    }
    \label{fig:shape_fraction}
\end{figure*}

Figure~\ref{fig:shape_fraction} (left) shows the class fractions for the fiducial mass-weighted shapes. In the low- and intermediate-mass bins, 94--98 percent of galaxies are classified as Spheroidal regardless of PI, and in the high-mass bin the galaxies are split between the Spheroidal and Disky classes. A $\chi^2$ test of independence finds no significant PI dependence in any bin ($p=0.29$, $0.11$, and $0.37$). The mass-weighted three-class scheme is therefore too coarse to recover the continuous $c/a$ trend.

We next repeat the classification using luminosity-weighted shapes, which more closely approximate the stellar populations emphasized by photometric observations (Figure~\ref{fig:shape_fraction}, right). Here the classes recover the environmental trend in the two lower-mass bins. In the low-mass bin, the Elongated fraction decreases from 28.4 percent for ${\rm PI}<1$ to 11.6 percent for ${\rm PI}>1$, while the Spheroidal fraction increases from 68.1 to 87.4 percent ($\chi^2=21.52$, $p=2.1\times10^{-5}$). The intermediate-mass bin shows a similar shift, with the Elongated fraction decreasing from 29.1 to 16.4 percent and the Spheroidal fraction increasing from 53.1 to 70.0 percent ($\chi^2=24.81$, $p=4.1\times10^{-6}$). The high-mass bin is dominated by Disky systems (73--89 percent) and shows no significant PI dependence ($p=0.10$).

Visual inspection of all 81 low-mass galaxies classified as Elongated in luminosity-weighted shape confirms that they are genuine stellar systems rather than numerical artifacts, whose elongation traces young stars in a few bright star-forming clumps or thin young disks embedded in a rounder, older stellar body.

The luminosity-weighted classification is more sensitive to the environmental trend in the lower-mass bins because luminosity weighting emphasizes younger stellar populations, which more often occupy thin or clumpy structures, whereas the mass-weighted shapes are dominated by the older, dynamically hotter component that is classified as Spheroidal in nearly all lower-mass galaxies. Broad morphology classes therefore capture the environmental dependence of shape only partially. In Section~\ref{sec:results_shape_sfh}, we investigate whether the trend in $c/a$ is connected to the formation-time trends presented in Section~\ref{sec:results_sfhs}.

\subsection{The interplay between environment, formation time, and shape}
\label{sec:results_shape_sfh}

In Sections~\ref{sec:results_sfhs}--\ref{sec:results_shape_classes}, we showed that PI is correlated with both formation time and galaxy shape. We now ask whether these two trends are connected: is the correlation between PI and shape accounted for by formation time, or does PI remain correlated with shape at fixed formation time? For this analysis, we compute $\tau_{90}$ from the same bound star particles within $r_\star^{80}$ of the stellar center that we use to measure shapes (Section~\ref{sec:shape_measurements}), so that formation time and shape describe the same stellar population. Because the shape reliability cut already requires at least 300 star particles, every galaxy with a reliable shape measurement has enough star particles in this aperture to measure $\tau_{90}$.

\begin{figure}
    \centering
    \includegraphics[width=1.0\linewidth, trim = 0 0 0 0]{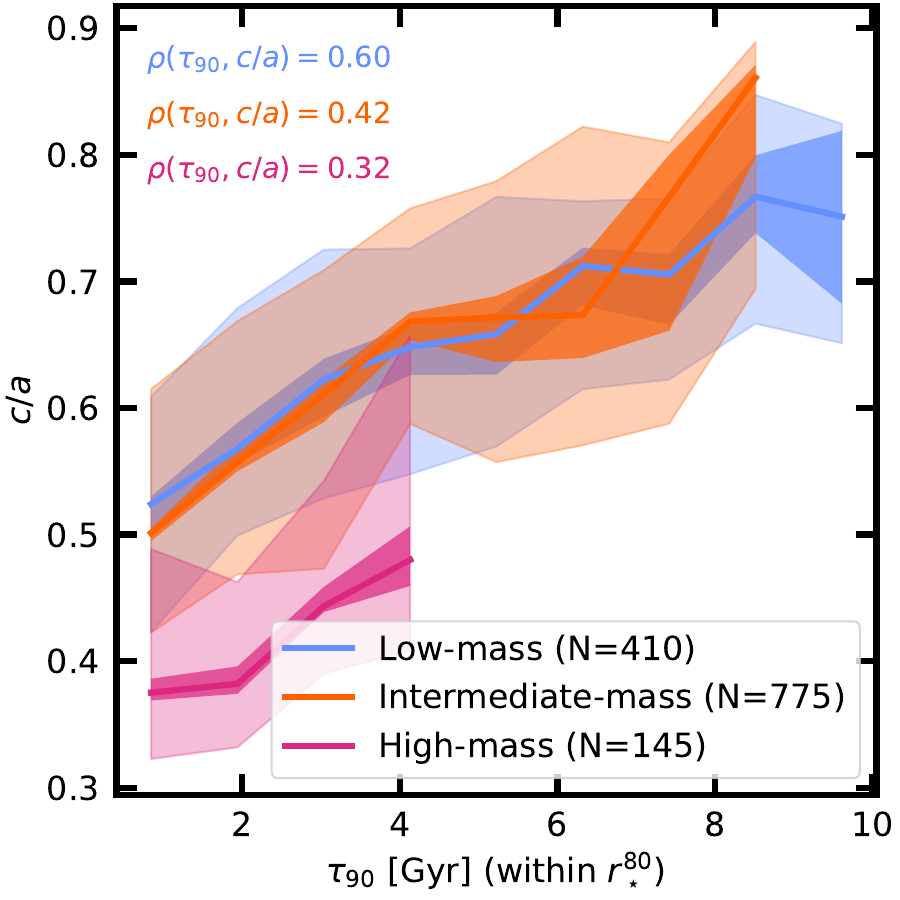}
    \caption{
    Minor-to-major axis ratio, $c/a$, as a function of aperture-matched $\tau_{90}$ for the low- (blue), intermediate- (orange), and high-mass (magenta) bins, both measured from the bound star particles within $r_\star^{80}$ of the stellar center. Lines show running medians, dark shaded bands show bootstrap 16th--84th percentile ranges of the median, and light shaded regions show the 16th--84th percentile ranges of the galaxy distribution. In all three bins, earlier-forming galaxies tend to have larger $c/a$; the high-mass bin spans a narrower range of $\tau_{90}$.
    }
    \label{fig:ca_vs_tau90}
\end{figure}

Figure~\ref{fig:ca_vs_tau90} shows $c/a$ as a function of this aperture-matched $\tau_{90}$. In all three bins, earlier-forming galaxies tend to have larger $c/a$ (Spearman correlations of $0.60$, $0.42$, and $0.32$ in the low-, intermediate-, and high-mass bins, as annotated in the figure, or $0.57$, $0.41$, and $0.37$ at fixed stellar mass; Table~\ref{tab:partial_correlation_summary}). In the low- and intermediate-mass bins, the median $c/a$ rises from $\sim0.5$ at $\tau_{90}\approx1\,{\rm Gyr}$ to $\sim0.75$--$0.85$ at $\tau_{90}\gtrsim8\,{\rm Gyr}$. Galaxies in the high-mass bin are flatter overall (smaller $c/a$) and span a narrower range of $\tau_{90}$, since few of them formed 90 percent of their stars more than $\sim4\,{\rm Gyr}$ ago; within that range, their median $c/a$ rises from $\sim0.37$ to $\sim0.48$.

Because PI, $\tau_{90}$, and $c/a$ are all correlated with one another, we use the partial Spearman correlations described in Section~\ref{sec:partial_correlation} to disentangle these relations. The key quantity is $\rho({\rm PI},c/a \mid \tau_{90},\log\mstar)$: if it is much smaller than $\rho({\rm PI},c/a \mid \log\mstar)$, then formation time accounts for most of the PI--shape relation.

\begin{figure*}
    \centering
    \includegraphics[width=1.0\textwidth, trim = 0 0 0 0]{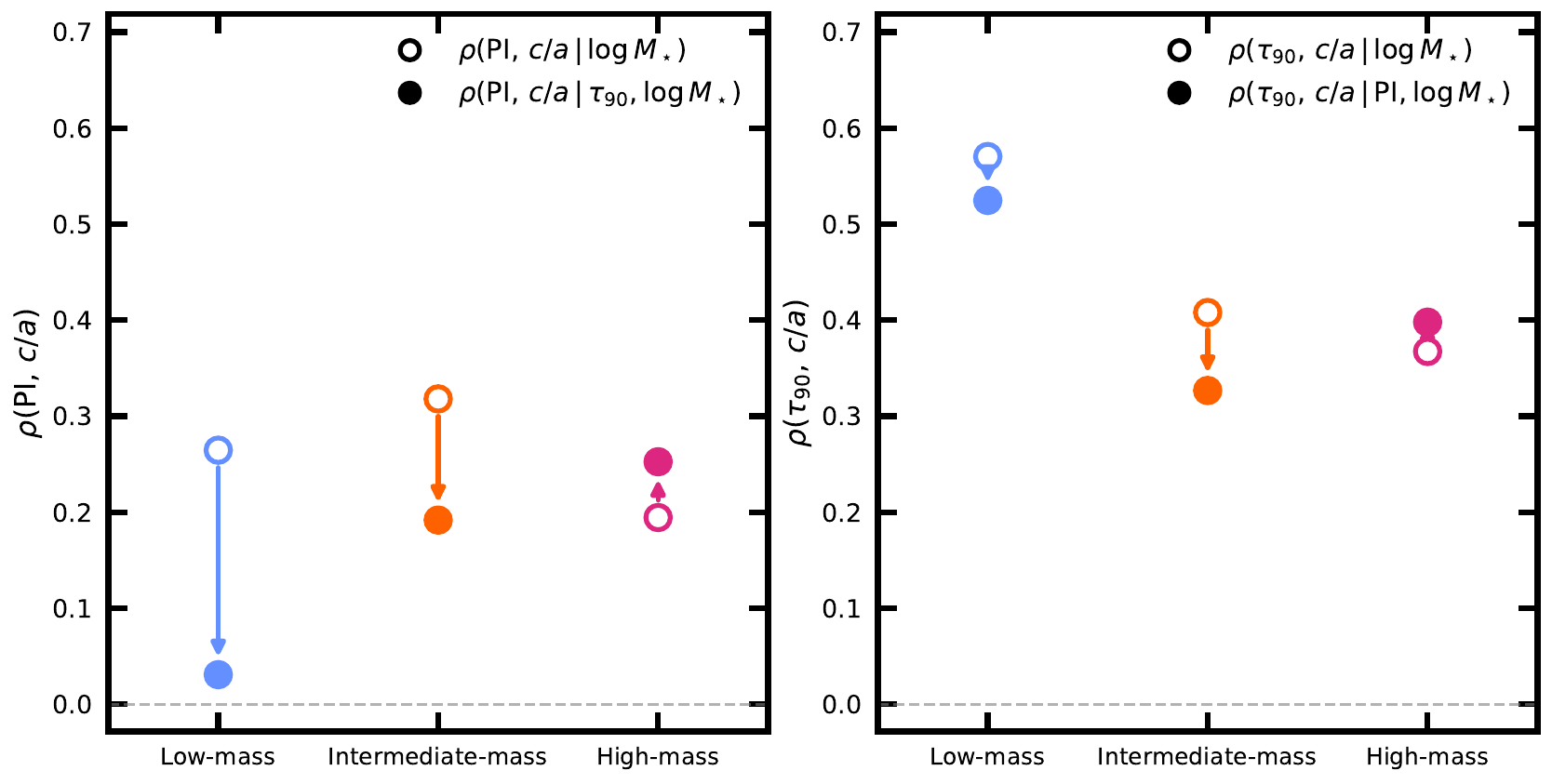}
    \caption{
    Spearman correlations at fixed stellar mass (open circles) and partial correlations that additionally control for a third variable (filled circles) for the fiducial mass-weighted shapes in the low- (blue), intermediate- (orange), and high-mass (magenta) bins; arrows point from the former to the latter; Table~\ref{tab:partial_correlation_summary} gives bootstrap uncertainties. \textbf{Left:} $\rho({\rm PI},c/a)$, with the partial correlation controlling for aperture-matched $\tau_{90}$. \textbf{Right:} $\rho(\tau_{90},c/a)$, with the partial correlation controlling for PI. Once formation time is controlled for, the PI--$c/a$ correlation becomes consistent with zero in the low-mass bin and is reduced in the intermediate-mass bin, but not in the high-mass bin, where PI is uncorrelated with $\tau_{90}$. The $\tau_{90}$--$c/a$ correlation remains strong in all three bins.
    }
    \label{fig:partial_correlation_dumbbell}
\end{figure*}

Table~\ref{tab:partial_correlation_summary} and Figure~\ref{fig:partial_correlation_dumbbell} show how each correlation changes when the controls are applied for the fiducial mass-weighted shapes, which we interpret following Section~\ref{sec:partial_correlation_interpretation} and Figure~\ref{fig:partial_correlation_schematic}. In the low-mass bin, at fixed stellar mass, PI is correlated with both $\tau_{90}$ ($\rho=0.42$) and $c/a$ ($\rho=0.26$), but $\tau_{90}$ is much more strongly correlated with $c/a$ ($\rho=0.57$). Additionally controlling for $\tau_{90}$ reduces the PI--$c/a$ correlation to a value consistent with zero ($\rho=0.03$, 95 percent bootstrap range $-0.06$ to $0.13$; $p=0.53$), while additionally controlling for PI leaves the $\tau_{90}$--$c/a$ correlation nearly intact ($0.57\rightarrow0.52$). For low-mass galaxies, formation time can therefore account for the correlation between environment and shape: at fixed formation time, high-PI galaxies do not have significantly larger $c/a$ than low-PI galaxies.

In the intermediate-mass bin, PI is again correlated with $\tau_{90}$ ($\rho=0.38$), and controlling for $\tau_{90}$ reduces the PI--$c/a$ correlation from $0.32$ to $0.19$, a decrease of about 40 percent, while controlling for PI reduces the $\tau_{90}$--$c/a$ correlation from $0.41$ to $0.33$. Formation time thus accounts for part of the PI--shape relation in this bin, and a modest partial correlation remains ($p=8\times10^{-8}$; 95 percent bootstrap range $0.12$--$0.26$). This residual is not driven by the most perturbed galaxies: it remains significant ($\rho=0.12$, $p=0.002$) when only galaxies with ${\rm PI}<10$ are included.

The high-mass bin differs in that PI is not significantly correlated with $\tau_{90}$ ($\rho=-0.11$, $p=0.20$), so formation time cannot account for the PI--$c/a$ correlation there. Controlling for $\tau_{90}$ leaves a positive partial correlation of $0.25$ ($p=0.002$; 95 percent bootstrap range $0.09$--$0.39$), while the $\tau_{90}$--$c/a$ correlation remains substantial ($0.37\rightarrow0.40$). This high-mass result rests on a small sample, and it is driven by the most perturbed galaxies: excluding galaxies with ${\rm PI}>100$, $>30$, or $>10$ reduces the partial correlation to $0.18$ ($p=0.04$), $0.13$ ($p=0.2$), and $0.03$ ($p=0.8$), respectively. We therefore regard the correlation between environment and shape at high mass as tentative.

\begin{table*}
    \centering
    \caption{
    Spearman correlations for the fiducial mass-weighted shapes in the low-, intermediate- (Int.), and high-mass bins. All correlations control for $\log\mstar$; arrows point from the mass-controlled correlation to the partial correlation that additionally controls for the third variable, and percentages give the change relative to the mass-controlled value. Brackets give 95 percent bootstrap ranges of the mass-controlled $\rho({\rm PI},\tau_{90} \mid \log\mstar)$ and of the partial correlations. $^\dagger$Consistent with zero ($p=0.53$); all other partial correlations have $p<0.01$.
    }
    \label{tab:partial_correlation_summary}
    \footnotesize
    \begin{tabular*}{\textwidth}{@{\extracolsep{\fill}}lccc}
        \hline
        Mass bin & $\rho({\rm PI},\tau_{90} \mid \log\mstar)$ & $\rho({\rm PI},c/a \mid \log\mstar)\rightarrow\rho({\rm PI},c/a \mid \tau_{90},\log\mstar)$ & $\rho(\tau_{90},c/a \mid \log\mstar)\rightarrow\rho(\tau_{90},c/a \mid {\rm PI},\log\mstar)$ \\
        \hline
        Low  & $0.42$ [$0.33$, $0.50$]   & $0.26\rightarrow0.03^\dagger$ [$-0.06$, $0.13$] ($-88\%$) & $0.57\rightarrow0.52$ [$0.45$, $0.59$] ($-8\%$) \\
        Int. & $0.38$ [$0.32$, $0.45$]   & $0.32\rightarrow0.19$ [$0.12$, $0.26$] ($-40\%$)          & $0.41\rightarrow0.33$ [$0.26$, $0.39$] ($-20\%$) \\
        High & $-0.11$ [$-0.27$, $0.07$] & $0.19\rightarrow0.25$ [$0.09$, $0.39$] ($+30\%$)          & $0.37\rightarrow0.40$ [$0.25$, $0.53$] ($+8\%$) \\
        \hline
    \end{tabular*}
\end{table*}

The luminosity-weighted shapes give qualitatively the same results: at fixed stellar mass, controlling for $\tau_{90}$ reduces the PI--$c/a$ correlation from $0.32$ to $0.06$ ($p=0.24$) in the low-mass bin and from $0.32$ to $0.19$ in the intermediate-mass bin, and increases it from $0.22$ to $0.30$ in the high-mass bin, while the $\tau_{90}$--$c/a$ correlation remains strong after controlling for PI ($0.60$, $0.37$, and $0.46$). We also verified that the change in the PI--$c/a$ partial correlation with stellar mass does not depend on the choice of bin edges by computing both partial correlations as a continuous function of stellar mass; the PI--$c/a$ partial correlation changes near $10^8\,\msun$, as in the binned analysis (Appendix~\ref{app:continuous_mass}).

Restricting the sample to central galaxies, whose halo radii and masses are defined differently from those of satellites (Section~\ref{sec:halo_finding}), gives the same mass-dependent pattern. In the low-mass bin, $\rho({\rm PI},c/a \mid \log\mstar)=0.19$ drops to $0.04$ ($p=0.46$) once $\tau_{90}$ is controlled for, and in all three bins the $\tau_{90}$--$c/a$ correlation remains strong after controlling for PI ($0.50$, $0.27$, and $0.36$). In the intermediate-mass bin, the residual PI--$c/a$ partial correlation of centrals ($0.09$, 95 percent bootstrap range $0.01$--$0.17$) is about half that of all galaxies, so satellites contribute part of it.

Overall, the partial-correlation analysis identifies two regimes. Earlier formation is correlated with larger $c/a$ at all stellar masses. Below $\sim10^8\,\msun$, the correlation between environment and shape is no longer significant once formation time is controlled for. Between $\sim10^8$ and $10^{10}\,\msun$, formation time accounts for part of this correlation, and a modest partial correlation between PI and $c/a$ remains at fixed formation time ($\rho\approx0.2$). For galaxies above $10^{10}\,\msun$, environment is not correlated with formation time (Figure~\ref{fig:tau_vs_pi}), and the residual PI--$c/a$ partial correlation is driven by the most perturbed systems. In Section~\ref{sec:results_halo_mass}, we examine how these correlations depend on halo mass.

\subsection{Environment and halo mass}\label{sec:results_halo_mass}

At fixed halo mass, the stellar masses of low-mass galaxies depend on the assembly histories of their halos \citep{Feldmann2019}, so at fixed stellar mass, halo mass may be correlated with both environment and formation time. We therefore examine how halo mass depends on PI at fixed stellar mass, and how the correlations of Section~\ref{sec:results_shape_sfh} change when halo mass is also held fixed.

\begin{figure}
    \centering
    \includegraphics[width=1.0\linewidth, trim = 0 0 0 0]{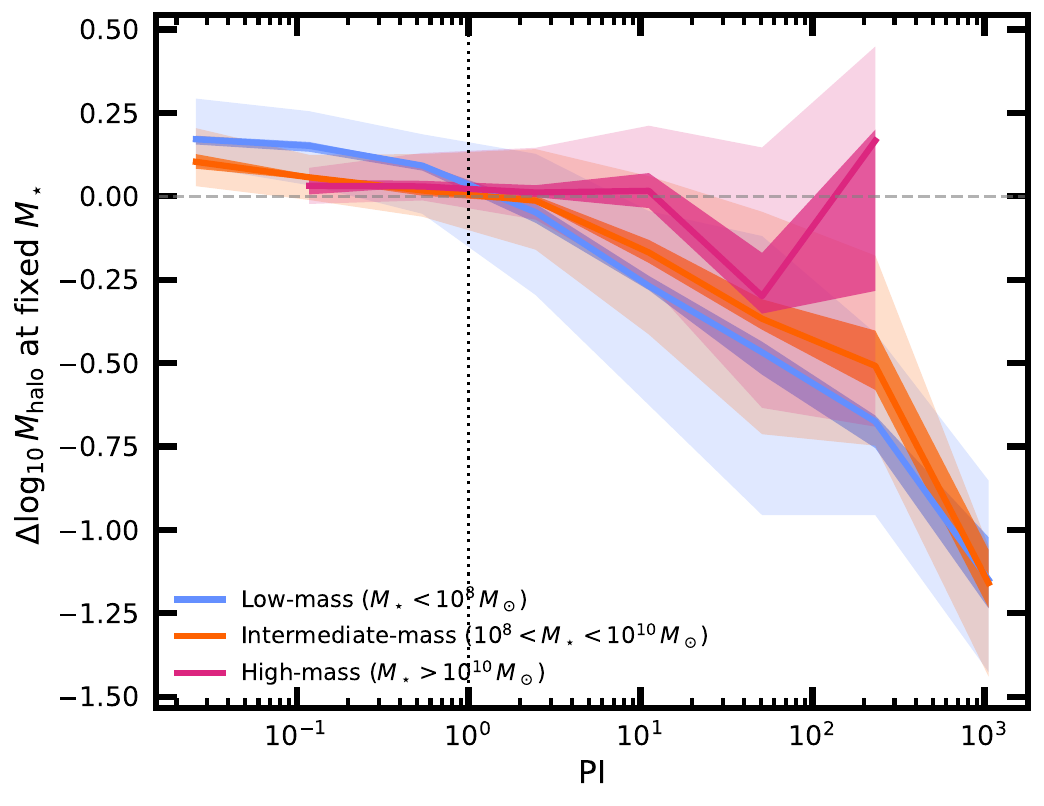}
    \caption{
    Offset of halo mass from the running median halo mass at fixed stellar mass, $\Delta\log M_{\rm halo}$, as a function of PI for the low- (blue), intermediate- (orange), and high-mass (magenta) bins (all 1720 galaxies). Lines show running medians, dark shaded bands show bootstrap 16th--84th percentile ranges of the median, and light shaded regions show the 16th--84th percentile ranges of the galaxy distribution. The dotted line marks ${\rm PI}=1$. In the low- and intermediate-mass bins, galaxies in stronger tidal environments have lower halo masses at fixed stellar mass.
    }
    \label{fig:halo_mass_offset}
\end{figure}

Figure~\ref{fig:halo_mass_offset} shows the offset of each galaxy's halo mass from the median halo mass at its stellar mass, $\Delta\log M_{\rm halo}$, as a function of PI. In the low- and intermediate-mass bins, galaxies in stronger tidal environments have lower halo masses at fixed stellar mass: the median $\Delta\log M_{\rm halo}$ is $0.12$ and $0.05$\,dex for low-PI galaxies and $-0.28$ and $-0.15$\,dex for high-PI galaxies, and it reaches $\approx-1$\,dex at ${\rm PI}\sim10^3$. The corresponding correlations are $\rho({\rm PI},\log M_{\rm halo} \mid \log\mstar)=-0.72$ and $-0.60$. Halo mass at fixed stellar mass is also correlated with formation time ($\rho(\tau_{90},\log M_{\rm halo} \mid \log\mstar)=-0.69$ and $-0.57$ for the whole-galaxy $\tau_{90}$). In the high-mass bin, $\Delta\log M_{\rm halo}$ shows no consistent trend with PI.

\begin{figure*}
    \centering
    \includegraphics[width=1.0\textwidth, trim = 0 0 0 0]{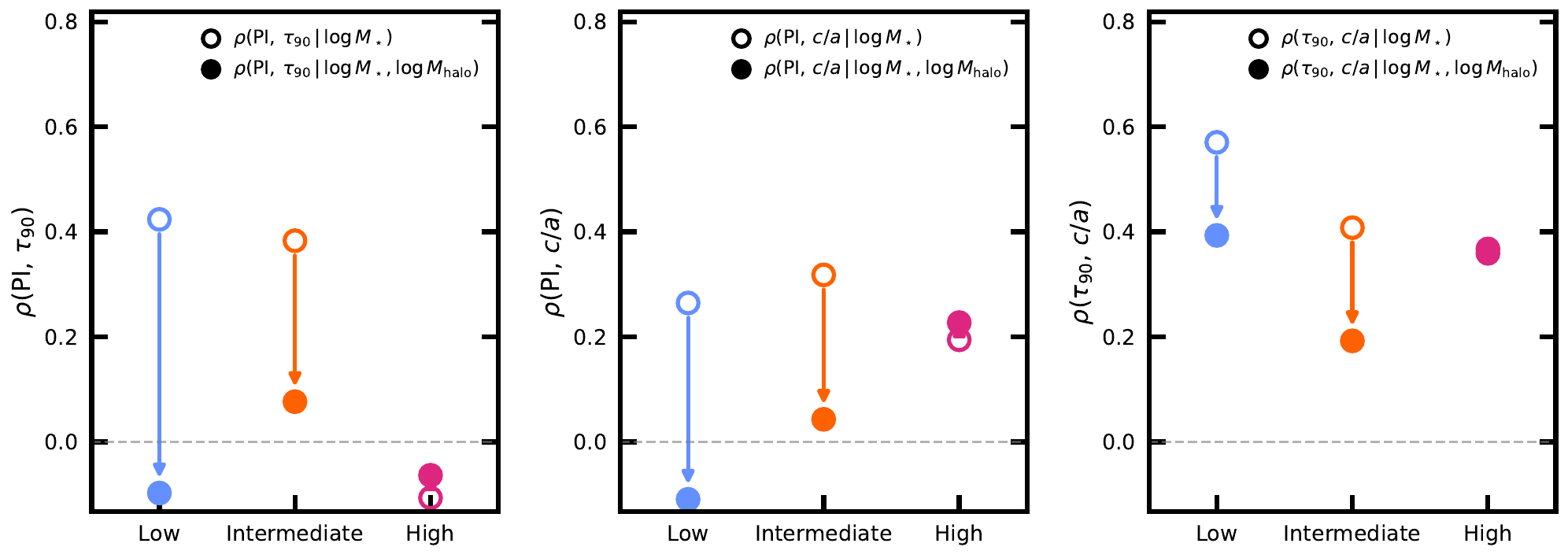}
    \caption{
    Spearman correlations at fixed stellar mass (open circles) and at fixed stellar mass and halo mass (filled circles) for the galaxies with reliable mass-weighted shapes in the low- (blue), intermediate- (orange), and high-mass (magenta) bins; arrows point from the former to the latter. \textbf{Left:} $\rho({\rm PI},\tau_{90})$; \textbf{middle:} $\rho({\rm PI},c/a)$; \textbf{right:} $\rho(\tau_{90},c/a)$, with aperture-matched $\tau_{90}$. In the low- and intermediate-mass bins, the correlations of PI with $\tau_{90}$ and $c/a$ become weak at fixed halo mass, while the $\tau_{90}$--$c/a$ correlation remains.
    }
    \label{fig:halo_mass_dumbbell}
\end{figure*}

Figure~\ref{fig:halo_mass_dumbbell} shows how the correlations among PI, $\tau_{90}$, and $c/a$ change when $\log M_{\rm halo}$ is controlled for in addition to $\log\mstar$. In the low- and intermediate-mass bins, the correlations of PI with $\tau_{90}$ ($0.42$ and $0.38$) and with $c/a$ ($0.26$ and $0.32$) are reduced to $|\rho|\lesssim0.1$ at fixed halo mass, and the PI--$c/a$ partial correlation at fixed $\tau_{90}$ and halo mass is $-0.08$ and $0.03$. In contrast, the $\tau_{90}$--$c/a$ correlation remains significant at fixed halo mass ($\rho=0.39$ and $0.19$, $p<10^{-7}$). In the high-mass bin, controlling for halo mass changes these correlations little.

For satellites, $M_{\rm halo}$ is the mass within the AHF truncation radius, which is itself set by the tidal field (Section~\ref{sec:halo_finding}), so part of the correlation between PI and halo mass could reflect the halo definition. For centrals, however, PI does not depend on the galaxy's own halo mass, because $\rhalo^3/\mhalo$ is the same for all virial halos at a given redshift (Section~\ref{sec:halo_finding} and Equation~\ref{eq:PI}). Centrals alone show the same behavior: at fixed stellar mass, $\rho({\rm PI},\log M_{\rm halo} \mid \log\mstar)=-0.57$ and $-0.32$ in the low- and intermediate-mass bins, and the correlations of PI with $\tau_{90}$ and $c/a$ have $|\rho|\leq0.12$ at fixed halo mass. We discuss the interpretation of these results in Section~\ref{sec:discussion_halo}.

\section{Discussion}\label{sec:discussion}

In FIREbox, environment is correlated with both the formation times and the shapes of galaxies, and formation time can account for most (below $\sim10^8\,\msun$) or part (between $\sim10^8$ and $10^{10}\,\msun$) of the correlation between environment and shape. We first place the formation-time trends in context (Section~\ref{sec:discussion_formation}), then discuss physical interpretations of the environment--shape connection (Section~\ref{sec:discussion_implications}), and finally summarize the caveats of our analysis (Section~\ref{sec:discussion_additional}).

\subsection{Formation times in context}\label{sec:discussion_formation}

Our formation-time results can be compared with the downsizing literature, which describes several observational trends, including the decline with cosmic time of the masses of actively star-forming galaxies \citep{Cowie1996} and the older stellar populations of more massive galaxies in the local Universe \citep{Thomas2005}; see \citet{Fontanot2009} for a review. Of these, the archaeological ages of present-day stellar populations are most directly comparable to our formation times, which also describe when the present-day stars formed. FIREbox follows this archaeological trend between our intermediate- and high-mass bins: at ${\rm PI}\lesssim10$, high-mass galaxies have larger $\tau_{30}$, $\tau_{50}$, and $\tau_{90}$ than intermediate-mass galaxies (Figure~\ref{fig:tau_vs_pi}). Our low-mass galaxies depart from it, forming a larger fraction of their present-day stellar mass at early times than either higher-mass bin, consistent with FIRE-2 zoom-in simulations of dwarf galaxies \citep{GarrisonKimmel2019b}; these galaxies lie below the stellar masses of the observational samples above, so this departure need not conflict with the downsizing trends. The absence of AGN feedback in FIREbox (Section~\ref{sec:discussion_additional}) may also affect the formation times of our high-mass galaxies, although we do not test this here. \citet{Thomas2005} also find that massive early-type galaxies in low-density environments are on average $\sim2$\,Gyr younger than those in high-density environments, qualitatively similar to the earlier formation we find in more strongly perturbing environments below $10^{10}\,\msun$.

Above $10^{10}\,\msun$, environment is no longer correlated with formation time, suggesting that the formation times of these galaxies, which sit in deeper potential wells, are set mainly by internal processes, such as stellar feedback and gas accretion. Using FIRE-2 zoom-in simulations, \citet{Santistevan2020} found that the main progenitors of Milky Way-mass galaxies in Local Group-like pairs emerged $\sim1.6$\,Gyr earlier than those of isolated hosts. At first glance, this result appears to conflict with ours, but the two analyses measure different quantities. Their metric traces when the main progenitor became dominant, at $z\sim3$--4, whereas our metrics trace when the bulk of the present-day stellar mass formed and are less sensitive to differences at the earliest epochs. The two results therefore need not be in tension, and together they are compatible with environment influencing the earliest assembly of massive galaxies more than the timing of their later star formation.

\subsection{Interpreting the environment--shape connection}\label{sec:discussion_implications}

Our results parallel an observational trend: at fixed stellar mass, environmental trends in galaxy color and star formation are stronger than those in morphology \citep{Bamford2009}. Similarly, in the low- and intermediate-mass bins, PI is more strongly correlated with $\tau_{90}$ ($\rho=0.42$ and $0.38$) than with $c/a$ ($\rho=0.26$ and $0.32$), with all values at fixed stellar mass (Table~\ref{tab:partial_correlation_summary}).

\subsubsection{Tidal transformation}\label{sec:discussion_tidal}

A natural interpretation of the connection between environment, formation time, and shape is that it reflects a population-level version of environmental transformation through tidal stirring. In tidal stirring models, repeated encounters with massive hosts or nearby structures remove angular momentum from stellar systems, heat stellar orbits, and can transform initially rotating dIrr galaxies into rounder, pressure-supported systems \citep[e.g.,][]{Mayer2001,Kazantzidis2011,Lokas2012,Lokas2015}. Environmental processing can also affect the gas supply and star formation activity of low-mass galaxies, providing a pathway by which structural transformation and earlier star formation truncation occur together \citep{Mayer2006,Weisz2011,Weisz2014,Fillingham2015,Fillingham2016,Digby2019,GarrisonKimmel2019b}. In this picture, environmental transformation should produce a coupled signature: galaxies in more strongly perturbing environments should stop forming stars earlier and develop dynamically hotter, rounder stellar distributions.

The statistical signature we find for low-mass FIREbox galaxies is qualitatively consistent with this expectation. These galaxies have earlier $\tau_{90}$ values in higher-PI environments, earlier-forming galaxies have larger $c/a$, and the PI--shape relation becomes consistent with zero once formation time is controlled for. The environmental shape trend is therefore closely tied to the environmental shift in formation time. In this sense, FIREbox provides a cosmological-volume counterpart to a transformation pathway that has often been studied in idealized or zoom-in settings \citep[e.g.,][]{Mayer2001,Kazantzidis2011,Lokas2012,Lokas2015}. \textit{However, the partial-correlation analysis is a correlational test and does not establish a causal pathway}, so it cannot by itself establish that tidal stirring is responsible for the observed trend.

Above $\sim10^8\,\msun$, the residual PI--shape partial correlation ($\rho\approx0.2$) indicates that the shapes of these galaxies may also reflect environmental processes that do not leave an imprint on $\tau_{90}$, for example through ongoing tidal heating or interactions with nearby companions; our analysis does not identify which. Between $10^8$ and $10^{10}\,\msun$, this residual persists among galaxies in only mildly perturbed environments (${\rm PI}<10$), but it is about half as large for centrals alone and consistent with zero at fixed halo mass (Sections~\ref{sec:results_shape_sfh} and~\ref{sec:results_halo_mass}).

Older stars in FIRE dwarf galaxies are more strongly heated by feedback-driven fluctuations regardless of environment \citep{ElBadry2016}, so the larger $c/a$ of high-PI galaxies could also arise if environment truncates late star formation and thereby removes a younger, flatter component, without heating the existing stars. To test this, we remeasured $c/a$ using only star particles older than 8\,Gyr within the same aperture, for galaxies with at least 300 such particles (Appendix~\ref{app:old_stars}). In the intermediate-mass bin, the old-star $c/a$ is as strongly correlated with PI at fixed stellar mass as the all-star $c/a$ ($\rho=0.31$ versus $0.32$), and this correlation persists at fixed $\tau_{90}$ ($\rho=0.24$, 95 percent bootstrap range $0.17$--$0.30$); stars older than 10\,Gyr give the same result. At intermediate mass, the environmental shape trend therefore extends to the old stellar population rather than arising only from the absence of young stars, consistent with environmental heating of existing stars. In the low-mass bin, where this test is limited to 215 of the more massive galaxies, the old-star $c/a$ is weakly correlated with PI ($\rho=0.16$, 95 percent range $0.03$--$0.28$), and, as for all stars, this correlation is consistent with zero at fixed $\tau_{90}$ ($\rho=-0.01$).

Above $10^{10}\,\msun$, galaxy shapes are still correlated with formation time, and the positive PI--shape correlation that remains is driven by the most perturbed systems (${\rm PI}\gtrsim30$), which reside in group environments. With only 145 galaxies in this bin, we regard this high-mass environmental signal as tentative.

\subsubsection{Halo stripping or early halo assembly}\label{sec:discussion_halo}

Similar statistical trends can, however, arise through pathways other than direct tidal transformation. Present-day environment may also trace earlier halo assembly or large-scale overdensity, meaning that PI may act both as a measure of the current tidal field and as a proxy for the density field in which galaxies formed. Galaxies in denser regions may therefore form earlier because their halos assembled earlier, or because their gas accretion histories differ from those of isolated systems. This interpretation is closely related to halo assembly bias, in which halos of similar mass can have different formation histories depending on their larger-scale environment \citep{Gao2005,Wechsler2006}. In this picture, the connection between environment, formation time, and structure may reflect a shared dependence on halo assembly history rather than direct tidal transformation alone.

The halo-mass results of Section~\ref{sec:results_halo_mass} provide a direct test of this possibility. At fixed stellar mass, low- and intermediate-mass galaxies in higher-PI environments have lower halo masses, consistent with the lower dark matter content of high-PI low-mass FIREbox galaxies found by \citet{Mercado2025}, and at fixed halo mass their correlations of PI with formation time and shape are weak ($|\rho|\lesssim0.1$). Our $z=0$ analysis cannot separate two interpretations of this result. If tidal stripping removes the outer dark matter halo and the gas supply of galaxies in strongly perturbing environments, halo mass lies on the environmental pathway, and controlling for it would remove the environmental signal by construction. Alternatively, halos that assembled early in dense regions may host galaxies with more stellar mass per unit halo mass and earlier formation times \citep{Feldmann2019}, in which case early halo assembly is a common origin of the environmental trends. The old-star result at intermediate mass (Section~\ref{sec:discussion_tidal}) favors the first interpretation, because tidal stripping of the halo would be accompanied by heating of the existing stars. Following the joint evolution of halo mass, stellar mass, and environment for individual galaxies would distinguish between these possibilities. In either case, the correlation between formation time and shape persists at fixed halo mass.

The distinction between direct transformation and a shared dependence on halo assembly is especially important for low-mass galaxies, whose stellar masses depend on halo assembly history even at fixed halo mass \citep{Feldmann2019}. These systems are most susceptible to environmental quenching, suppressed late-time gas accretion, and preprocessing in overdense regions. Their earlier $\tau_{90}$ values could therefore arise because dense environments remove gas or prevent its replenishment at late times, cutting off the fuel needed to sustain star formation to the present day, rather than because repeated tidal encounters directly heat the stellar body. In this case, the correlation between earlier formation and larger $c/a$ could reflect the longer time available for cumulative dynamical evolution, or a distinct structural pathway for early-forming systems, whose stars may be born in thicker configurations, as found in Milky Way-mass FIRE-2 simulations \citep{Yu2021,Yu2023}, rather than tidal heating by the specific neighbors that dominate PI at $z=0$.

\subsubsection{Synthesis and observational prediction}\label{sec:discussion_synthesis}

Overall, our results suggest a mass-dependent picture: environment is correlated with the shapes of low-mass galaxies largely through its correlation with their formation times, while between $10^8$ and $10^{10}\,\msun$ an additional, modest correlation between environment and shape remains at fixed formation time, though not at fixed halo mass. Below $10^{10}\,\msun$, these environmental trends are closely tied to halo mass at fixed stellar mass, whereas the connection between formation time and shape persists at fixed halo mass. Distinguishing among the physical interpretations above will require analyses that connect present-day environment to time-dependent environmental histories, halo assembly histories, gas accretion histories, and the dynamical evolution of stellar shapes.

These results lead to an observational prediction: \textit{at fixed stellar mass, dwarf galaxies in stronger tidal environments should appear rounder in projection, with the difference largely tracking the ages of their stellar populations.} Our luminosity-weighted shapes, which more closely approximate observed light, show the same environmental trend (Section~\ref{sec:results_shape_sfh}). Photometric studies of the projected and intrinsic shapes of dwarf galaxies provide a natural test \citep{KadoFong2020,Carlsten2021}; for example, \citet{Carlsten2021} find that early-type dwarf satellites are rounder than late-type dwarfs at fixed luminosity, qualitatively consistent with this picture. Comparing axis-ratio distributions of dwarfs in different environments at fixed stellar mass and stellar age would test whether environment is correlated with shape beyond the effect of formation time.

\subsection{Caveats and limitations}\label{sec:discussion_additional}

While our analysis connects environment, formation time, and intrinsic stellar structure within a cosmological population, several additional physical processes and observational effects remain unexplored. One important caveat is that FIREbox does not include feedback from active galactic nuclei (AGN), which is expected to become increasingly important toward the high-mass end of our sample. AGN feedback is thought to play a key role in regulating the evolution of massive galaxies by suppressing star formation, altering gas accretion, and influencing the transition between disk-dominated and spheroid-dominated systems. Previous simulation studies show that AGN feedback can strongly affect the structural evolution of massive galaxies, producing quenched, spheroid-dominated systems and preventing the overly compact, star-forming galaxies that often emerge in simulations without AGN feedback \citep[e.g.,][]{Taylor2015,Dubois2016}. Analyses of FIRE-3 simulations that incorporate AGN feedback similarly find that AGN activity can modify the stellar populations and structural properties of massive galaxies, producing older stellar populations, quenched systems, and galaxy scaling relations that better match observations compared with simulations without AGN feedback \citep{Byrne2024}. \citet{Fofie2026} further use FIRE-2 zoom simulations to investigate how different AGN feedback implementations influence the morphology and dynamical structure of Milky Way-mass galaxies. Without AGN feedback, our high-mass galaxies may continue forming stars, and maintaining disks, later than they would otherwise, which would bias their formation times late and their $c/a$ values low; we do not test this here. Consequently, our results at high stellar mass should be interpreted as describing the connection between environment, formation time, and structure in galaxies whose evolution is not dominated by AGN-driven transformation. Future comparisons with simulations that include AGN feedback will therefore be essential for determining how the relationship between environment, formation time, and morphology changes once AGN-driven evolution is incorporated.

Another caveat concerns the earliest star formation in our low-mass galaxies. FIREbox adopts the ultraviolet background of \citet{FaucherGiguere2009}, which reionizes the universe at $z\approx10$, earlier than the $z\sim8$ favored by more recent empirical constraints \citep{FaucherGiguere2020,Wetzel2023}. Because reionization heats the intergalactic gas and can suppress star formation in low-mass halos, this earlier reionization may reduce the stellar mass formed at lookback times $\gtrsim13$\,Gyr in our lowest-mass galaxies. This would affect $\tau_{30}$ more than $\tau_{50}$ and $\tau_{90}$, which trace later star formation and carry our strongest environmental signal. FIRE-2 resimulations with a later-reionizing background \citep{Wetzel2025} provide a way to test this sensitivity.

Our shapes may also be affected by spurious collisional heating of star particles by the more massive dark matter particles, which artificially increases stellar velocity dispersions and disk thicknesses, particularly in halos resolved with relatively few dark matter particles \citep{Ludlow2021,Ludlow2023}. Because this heating accumulates over time, older stellar populations have been exposed to it for longer, which could contribute to the correlation between $\tau_{90}$ and $c/a$, especially in our low-mass bin. Testing this requires comparing simulations at different resolution, such as lower-resolution FIREbox runs or higher-resolution FIRE-2 zoom-in simulations, which we leave to future work.

A final consideration is that our analysis focuses on the intrinsic three-dimensional stellar distributions of simulated galaxies rather than mock observations. This approach allows us to directly quantify the underlying stellar mass distribution and its connection to galaxy formation times without introducing additional effects associated with projection, dust attenuation, stellar population modeling, or observational measurement uncertainties. However, observations typically infer galaxy structure from projected light distributions and star formation histories from integrated photometric or spectroscopic measurements. Constructing mock observations and recovering galaxy properties using observational analysis techniques will therefore provide an important extension by connecting the intrinsic trends identified here to quantities that can be directly compared with surveys. Previous work has demonstrated the value of this approach by generating realistic mock observations of FIREbox galaxies and applying observationally motivated morphology measurements to characterize the relationship between intrinsic and observable galaxy properties \citep{Klein2026,Geng2026}. In this work, we partially bridge this gap by comparing mass-weighted and luminosity-weighted stellar shapes. The luminosity-weighted measurements emphasize younger stellar populations and more closely approximate the stellar components probed by observations, while recovering the same qualitative environmental trends found in the intrinsic shapes. However, dust could dim the young, star-forming clumps that drive some of the luminosity-weighted trends, so these measurements should be compared with observations with caution. Future analyses incorporating full mock imaging, photometry, and observationally inferred star formation histories will be valuable for determining how the environmental connections identified here manifest in observed galaxy populations.

\section{Summary and Conclusions}\label{sec:conclusion}

In this work, we investigate how galaxy environment, formation time, and intrinsic stellar structure are connected using the FIREbox cosmological volume simulation. Unlike many previous studies that examine these quantities separately, we analyze their connection within a single cosmological population. We quantify environment using a continuous Perturbation Index (PI), measure formation times from cumulative archaeological star formation histories, and characterize galaxy structure using intrinsic three-dimensional stellar shapes. Our main conclusions are:

\begin{enumerate}

\item \textbf{Formation times are most strongly correlated with stellar mass, but environment plays a secondary role in the low- and intermediate-mass bins.}  
Galaxies in the low-mass bin ($\mstar<10^8\,\msun$) form their stars earlier on average than those in the intermediate- ($10^8$--$10^{10}\,\msun$) and high-mass ($\mstar>10^{10}\,\msun$) bins, demonstrating the strong dependence of formation time on galaxy mass. However, within the low- and intermediate-mass bins, galaxies in more strongly perturbing environments tend to have earlier formation times, with the strongest environmental dependence appearing in $\tau_{90}$ for the low-mass galaxies. This indicates that the present-day environment is most strongly correlated with the late-time star formation of low-mass systems.

\item \textbf{Environmental differences in formation time are strongest in the inner regions of low-mass galaxies.}  
Radially resolved formation times show that the environmental dependence of formation time is most apparent for stars currently located within the inner stellar half-mass radius of low-mass galaxies. High-PI low-mass galaxies contain older inner stellar populations than low-PI systems of comparable mass, while environmental differences become weaker at larger present-day radii. This suggests that the environmental dependence of formation time is not uniformly distributed throughout the stellar component.

\item \textbf{Galaxy structure exhibits continuous environmental trends that are not fully captured by traditional morphology classifications.}  
Within each stellar-mass bin, galaxies in stronger tidal environments have larger minor-to-major axis ratios ($c/a$), indicating more spherical stellar distributions in the low- and intermediate-mass bins, where $b/a$ also increases, and stellar distributions that are thicker along the minor axis in the high-mass bin. This trend is clearest in the continuous axis ratios and is lost when galaxies are grouped into broad morphology classes based on their mass-weighted shapes, because nearly all galaxies below $10^{10}\,\msun$ are then classified as Spheroidal regardless of environment. Classifying galaxies by their luminosity-weighted shapes instead partly recovers the trend: below $10^{10}\,\msun$, galaxies in more strongly perturbing environments are less often classified as Elongated and more often as Spheroidal.

\item \textbf{Formation time links environment and shape in a mass-dependent way.}  
Earlier-forming galaxies have larger $c/a$ at all stellar masses. Below $10^8\,\msun$, controlling for $\tau_{90}$ and stellar mass reduces the PI--$c/a$ correlation to a value consistent with zero. Between $10^8$ and $10^{10}\,\msun$, it reduces this correlation, leaving a modest residual partial correlation ($\rho\approx0.2$). Above $10^{10}\,\msun$, environment is not correlated with formation time, and the residual PI--shape partial correlation is driven by the most perturbed systems and is tentative. In other words, low-mass galaxies in stronger tidal environments are more spherical to the extent expected from their earlier formation, while at higher masses environment remains correlated with shape at fixed formation time. A continuous analysis in stellar mass confirms that these regimes do not depend on the choice of bin edges.

\item \textbf{Below $10^{10}\,\msun$, the environmental trends are tied to halo mass at fixed stellar mass.}  
At fixed stellar mass, galaxies in stronger tidal environments have lower halo masses, and at fixed halo mass, PI is only weakly correlated with formation time and shape ($|\rho|\lesssim0.1$), for all galaxies and for centrals alone. Formation time and shape remain correlated at fixed halo mass.

\item \textbf{The relationship between environment, formation time, and structure likely reflects multiple physical mechanisms.}  
Our results are consistent with a population-level connection between environmental transformation and structural evolution, particularly for low-mass galaxies where tidal interactions and environmental regulation of gas supply are expected to be important. However, the halo-mass dependence of these trends is consistent both with tidal stripping, in which halo mass lies on the environmental pathway, and with early halo assembly in dense regions as a common origin of the environmental trends. Combining time-dependent environmental histories and halo assembly histories will be necessary to distinguish between direct environmental transformation and shared dependence on galaxy formation histories.

\end{enumerate}

These results demonstrate that galaxy environment is connected to both when galaxies formed their stars and how their present-day stellar structures are organized. By connecting continuous environmental measures, cumulative star formation histories, and intrinsic stellar shapes within the same cosmological population, our work provides a framework for understanding how internal galaxy processes and external environment are connected across galaxy populations.

\section*{Acknowledgments}
FJM was supported in part by the National Science Foundation (NSF) Math and Physical Sciences (MPS) Award AST-2316748 during the completion of this work. JM is funded by a Pomona College Large Research Grant.

\software{The functionalities provided by the following python packages played a critical role in the analysis and visualizations presented in this paper: \textbf{\textsc{matplotlib}} \citep{Hunter2007}, \textbf{\textsc{NumPy}} \citep{Harris2020}, \textbf{\textsc{SciPy}} \citep{Virtanen2020}, \textbf{\textsc{pandas}} \citep{McKinney2010}, \textbf{\textsc{h5py}} \citep{Collette2013}, and \textbf{\textsc{Astropy}} \citep{AstropyCollaboration2013,AstropyCollaboration2018,AstropyCollaboration2022}. Halo and galaxy catalogs were constructed with the AMIGA Halo Finder (\textbf{\textsc{AHF}}; \citealt{Knollmann2009}).}

\appendix

\section{Partial correlations as a continuous function of stellar mass}\label{app:continuous_mass}
\restartappendixnumbering

\begin{figure*}
    \centering
    \includegraphics[width=1.0\textwidth, trim = 0 0 0 0]{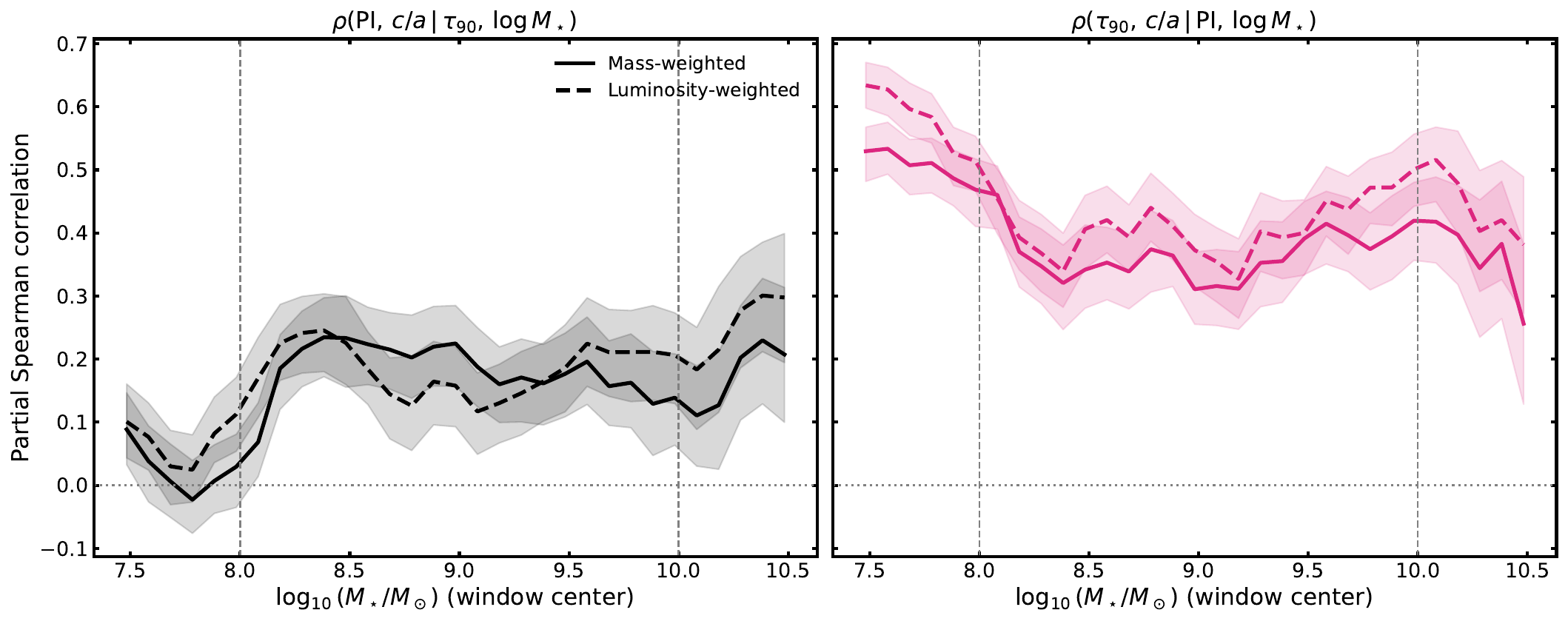}
    \caption{
    Partial Spearman correlations as a continuous function of stellar mass for mass-weighted (solid lines) and luminosity-weighted (dashed lines) shapes; shaded regions show bootstrap 16th--84th percentile ranges. \textbf{Left:} $\rho({\rm PI},c/a \mid \tau_{90},\log\mstar)$; \textbf{right:} $\rho(\tau_{90},c/a \mid {\rm PI},\log\mstar)$. Each point uses the galaxies with reliable shapes in a 0.6\,dex window of $\log\mstar$ centered on the plotted value, stepped by 0.1\,dex. Vertical dashed lines mark the bin edges at $10^8$ and $10^{10}\,\msun$. The $\tau_{90}$--$c/a$ partial correlation is positive at all masses, while the PI--$c/a$ partial correlation is nonzero only above $\sim10^{8.2}\,\msun$.
    }
    \label{fig:partial_correlation_vs_mass}
\end{figure*}

To test whether the results of Section~\ref{sec:results_shape_sfh} depend on where the stellar-mass bin edges are placed, Figure~\ref{fig:partial_correlation_vs_mass} shows both partial correlations as a continuous function of stellar mass, computed in overlapping windows of width 0.6\,dex with $\log\mstar$ still controlled for within each window. The PI--$c/a$ partial correlation is consistent with zero for $\mstar\lesssim10^{8.1}\,\msun$ and rises to $\approx0.15$--$0.25$ above $\sim10^{8.2}\,\msun$, remaining at that level to the highest masses probed. The $\tau_{90}$--$c/a$ partial correlation is positive at all masses, from $\approx0.5$--$0.65$ at the low-mass end to $\approx0.3$--$0.5$ above $10^{8.2}\,\msun$. Both weighting schemes show the same behavior.

\section{Shapes of old stars}\label{app:old_stars}
\restartappendixnumbering

\begin{figure}
    \centering
    \includegraphics[width=1.0\linewidth, trim = 0 0 0 0]{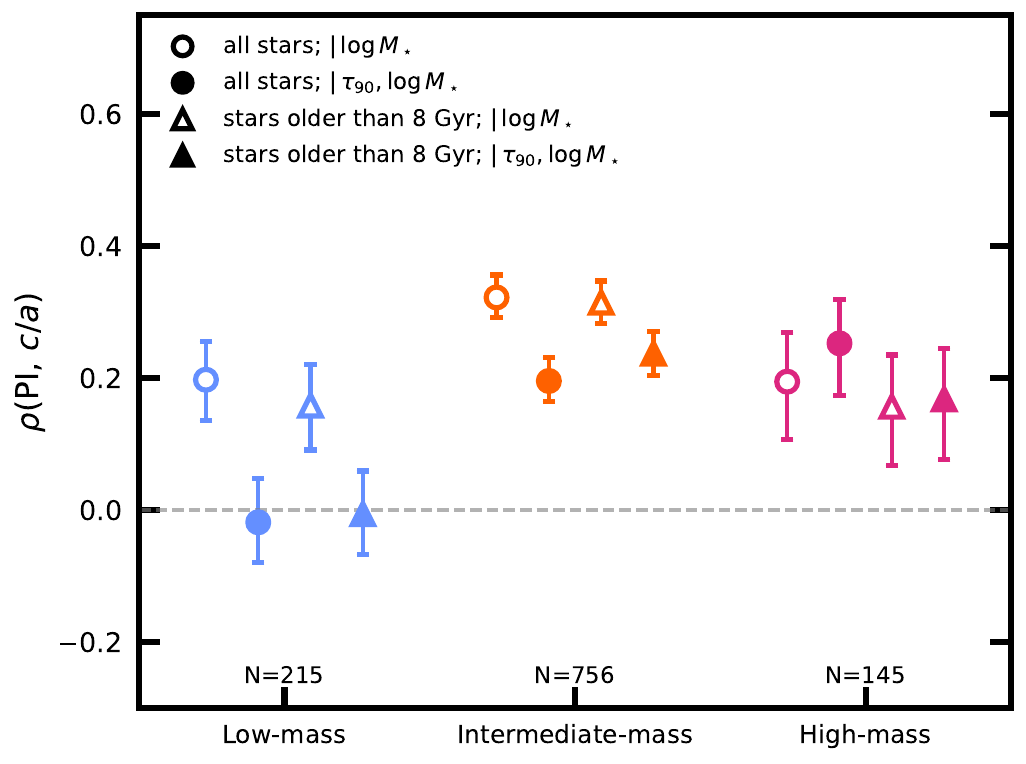}
    \caption{
    $\rho({\rm PI},c/a)$ at fixed $\log\mstar$ (open symbols) and at fixed $\tau_{90}$ and $\log\mstar$ (filled symbols) for the $c/a$ of all stars (circles) and of stars older than 8\,Gyr (triangles), for the same galaxies, which have at least 300 old star particles; numbers give the sample size in each bin. Error bars show bootstrap 16th--84th percentile ranges. In the intermediate-mass bin, the old-star $c/a$ remains correlated with PI at fixed $\tau_{90}$.
    }
    \label{fig:old_star_shapes}
\end{figure}

To test whether the environmental shape trend arises only from the absence of young, flatter components in high-PI galaxies (Section~\ref{sec:discussion_tidal}), we remeasure $c/a$ using only star particles older than 8\,Gyr. We use the same stars, stellar center, and $r_\star^{80}$ aperture as for the fiducial shapes and keep only galaxies with at least 300 old star particles, which leaves 215, 756, and 145 galaxies in the low-, intermediate-, and high-mass bins; the low-mass subsample is restricted to the more massive galaxies of that bin. Figure~\ref{fig:old_star_shapes} compares $\rho({\rm PI},c/a)$ for the all-star and old-star shapes of the same galaxies, at fixed stellar mass and at fixed $\tau_{90}$ and stellar mass. Using stars older than 10\,Gyr instead gives the same result in the intermediate-mass bin ($\rho=0.33$ at fixed stellar mass), while the low-mass subsample becomes too small (79 galaxies) for a significant measurement.

\bibliography{refs}{}

\begin{thebibliography}{}
\expandafter\ifx\csname natexlab\endcsname\relax\def\natexlab#1{#1}\fi
\providecommand{\url}[1]{\href{#1}{#1}}
\providecommand{\dodoi}[1]{doi:~\href{http://doi.org/#1}{\nolinkurl{#1}}}
\providecommand{\doeprint}[1]{\href{http://ascl.net/#1}{\nolinkurl{http://ascl.net/#1}}}
\providecommand{\doarXiv}[1]{\href{https://arxiv.org/abs/#1}{\nolinkurl{https://arxiv.org/abs/#1}}}

\bibitem[{{Allgood} {et~al.}(2006){Allgood}, {Flores}, {Primack}, {Kravtsov},
  {Wechsler}, {Faltenbacher}, \& {Bullock}}]{Allgood2006}
{Allgood}, B., {Flores}, R.~A., {Primack}, J.~R., {et~al.} 2006, \mnras, 367,
  1781, \dodoi{10.1111/j.1365-2966.2006.10094.x}

\bibitem[{{Arora} {et~al.}(2025){Arora}, {Garavito-Camargo}, {Sanderson},
  {Weinberg}, {Petersen}, {Varela-Lavin}, {G{\'o}mez}, {Johnston}, {Laporte},
  {Shipp}, {Hunt}, {Besla}, {Darragh-Ford}, {Panithanpaisal}, \&
  {Daniel}}]{Arora2025}
{Arora}, A., {Garavito-Camargo}, N., {Sanderson}, R.~E., {et~al.} 2025, \apj,
  988, 190, \dodoi{10.3847/1538-4357/ade30d}

\bibitem[{{Astropy Collaboration}(2022)}]{AstropyCollaboration2022}
{Astropy Collaboration}. 2022, \apj, 935, 167, \dodoi{10.3847/1538-4357/ac7c74}

\bibitem[{{Astropy Collaboration} {et~al.}(2018){Astropy Collaboration},
  {Price-Whelan}, {Sip{\H{o}}cz}, {G{\"u}nther}, {Lim}, {Crawford},
  {et~al.}}]{AstropyCollaboration2018}
{Astropy Collaboration}, {Price-Whelan}, A.~M., {Sip{\H{o}}cz}, B.~M., {et~al.}
  2018, \aj, 156, 123, \dodoi{10.3847/1538-3881/aabc4f}

\bibitem[{{Astropy Collaboration} {et~al.}(2013){Astropy Collaboration},
  {Robitaille}, {Tollerud}, {Greenfield}, {Droettboom}, {Bray},
  {et~al.}}]{AstropyCollaboration2013}
{Astropy Collaboration}, {Robitaille}, T.~P., {Tollerud}, E.~J., {et~al.} 2013,
  \aap, 558, A33, \dodoi{10.1051/0004-6361/201322068}

\bibitem[{{Baba} {et~al.}(2004){Baba}, {Shibata}, \& {Sibuya}}]{Baba2004}
{Baba}, K., {Shibata}, R., \& {Sibuya}, M. 2004, Australian \& New Zealand
  Journal of Statistics, 46, 657, \dodoi{10.1111/j.1467-842x.2004.00360.x}

\bibitem[{{Bailin} \& {Steinmetz}(2005)}]{BailinSteinmetz2005}
{Bailin}, J., \& {Steinmetz}, M. 2005, \apj, 627, 647, \dodoi{10.1086/430397}

\bibitem[{{Bait} {et~al.}(2017){Bait}, {Barway}, \& {Wadadekar}}]{Bait2017}
{Bait}, O., {Barway}, S., \& {Wadadekar}, Y. 2017, \mnras, 471, 2687,
  \dodoi{10.1093/mnras/stx1688}

\bibitem[{{Bamford} {et~al.}(2009){Bamford}, {Nichol}, {Baldry}, {Land},
  {Lintott}, {Schawinski}, {Slosar}, {Szalay}, {Thomas}, {Torki}, {Andreescu},
  {Edmondson}, {Miller}, {Murray}, {Raddick}, \& {Vandenberg}}]{Bamford2009}
{Bamford}, S.~P., {Nichol}, R.~C., {Baldry}, I.~K., {et~al.} 2009, \mnras, 393,
  1324, \dodoi{10.1111/j.1365-2966.2008.14252.x}

\bibitem[{{Bellardini} {et~al.}(2026){Bellardini}, {McCluskey}, {Wetzel},
  {Steel}, {Loebman}, \& {Moreno}}]{Bellardini2026}
{Bellardini}, M.~A., {McCluskey}, F., {Wetzel}, A., {et~al.} 2026, arXiv
  e-prints, arXiv:2605.06781

\bibitem[{{Bellardini} {et~al.}(2022){Bellardini}, {Wetzel}, {Loebman}, \&
  {Bailin}}]{Bellardini2022}
{Bellardini}, M.~A., {Wetzel}, A., {Loebman}, S.~R., \& {Bailin}, J. 2022,
  \mnras, 514, 4270, \dodoi{10.1093/mnras/stac1637}

\bibitem[{{Bellardini} {et~al.}(2021){Bellardini}, {Wetzel}, {Loebman},
  {Faucher-Gigu{\`e}re}, {Ma}, \& {Feldmann}}]{Bellardini2021}
{Bellardini}, M.~A., {Wetzel}, A., {Loebman}, S.~R., {et~al.} 2021, \mnras,
  505, 4586, \dodoi{10.1093/mnras/stab1606}

\bibitem[{{Benavides} {et~al.}(2025){Benavides}, {Sales}, {Wetzel}, {Moreno},
  {Feldmann}, {Mercado}, {Bullock}, {Hopkins}, {Faucher-Gigu{\'e}re}, {Stern},
  {Wheeler}, \& {Kere{\v{s}}}}]{Benavides2025}
{Benavides}, J.~A., {Sales}, L.~V., {Wetzel}, A., {et~al.} 2025, \mnras,
  \dodoi{10.1093/mnras/staf1847}

\bibitem[{{Benincasa} {et~al.}(2020){Benincasa}, {Loebman}, {Wetzel},
  {Hopkins}, {Murray}, {Bellardini}, {Faucher-Gigu{\`e}re}, {Guszejnov}, \&
  {Orr}}]{Benincasa2020}
{Benincasa}, S.~M., {Loebman}, S.~R., {Wetzel}, A., {et~al.} 2020, \mnras, 497,
  3993, \dodoi{10.1093/mnras/staa2116}

\bibitem[{{Blanton} \& {Moustakas}(2009)}]{Blanton2009}
{Blanton}, M.~R., \& {Moustakas}, J. 2009, \araa, 47, 159,
  \dodoi{10.1146/annurev-astro-082708-101734}

\bibitem[{{Bluck} {et~al.}(2020){Bluck}, {Maiolino}, {S{\'a}nchez}, {Ellison},
  {Thorp}, {Piotrowska}, {Teimoorinia}, \& {Bundy}}]{Bluck2020}
{Bluck}, A. F.~L., {Maiolino}, R., {S{\'a}nchez}, S.~F., {et~al.} 2020, \mnras,
  492, 96, \dodoi{10.1093/mnras/stz3264}

\bibitem[{{Bluck} {et~al.}(2019){Bluck}, {Bottrell}, {Teimoorinia},
  {Henriques}, {Mendel}, {Ellison}, {Thanjavur}, {Simard}, {Patton},
  {Conselice}, {Moreno}, \& {Woo}}]{Bluck2019}
{Bluck}, A. F.~L., {Bottrell}, C., {Teimoorinia}, H., {et~al.} 2019, \mnras,
  485, 666, \dodoi{10.1093/mnras/stz363}

\bibitem[{{Boselli} \& {Gavazzi}(2006)}]{Boselli2006}
{Boselli}, A., \& {Gavazzi}, G. 2006, \pasp, 118, 517, \dodoi{10.1086/500691}

\bibitem[{{Bryan} \& {Norman}(1998)}]{Bryan1998}
{Bryan}, G.~L., \& {Norman}, M.~L. 1998, \apj, 495, 80, \dodoi{10.1086/305262}

\bibitem[{{Byrne} {et~al.}(2024){Byrne}, {Faucher-Gigu{\`e}re}, {Wellons},
  {Hopkins}, {Angl{\'e}s-Alc{\'a}zar}, {Sultan}, {Wijers}, {Moreno}, \&
  {Ponnada}}]{Byrne2024}
{Byrne}, L., {Faucher-Gigu{\`e}re}, C.-A., {Wellons}, S., {et~al.} 2024, \apj,
  973, 149, \dodoi{10.3847/1538-4357/ad67ca}

\bibitem[{{Carlsten} {et~al.}(2021){Carlsten}, {Greene}, {Greco}, {Beaton}, \&
  {Kado-Fong}}]{Carlsten2021}
{Carlsten}, S.~G., {Greene}, J.~E., {Greco}, J.~P., {Beaton}, R.~L., \&
  {Kado-Fong}, E. 2021, \apj, 922, 267, \dodoi{10.3847/1538-4357/ac2581}

\bibitem[{{Chabrier}(2003)}]{Chabrier2003}
{Chabrier}, G. 2003, \pasp, 115, 763, \dodoi{10.1086/376392}

\bibitem[{{Christensen} {et~al.}(2024){Christensen}, {Brooks}, {Munshi},
  {Riggs}, {Van Nest}, {Akins}, {Quinn}, \& {Chamberland}}]{Christensen2024}
{Christensen}, C.~R., {Brooks}, A.~M., {Munshi}, F., {et~al.} 2024, \apj, 961,
  236, \dodoi{10.3847/1538-4357/ad0c5a}

\bibitem[{{Chua} {et~al.}(2019){Chua}, {Pillepich}, {Vogelsberger}, \&
  {Hernquist}}]{Chua2019}
{Chua}, K. T.~E., {Pillepich}, A., {Vogelsberger}, M., \& {Hernquist}, L. 2019,
  \mnras, 484, 476, \dodoi{10.1093/mnras/sty3531}

\bibitem[{{Collette}(2013)}]{Collette2013}
{Collette}, A. 2013, {Python and HDF5: Unlocking Scientific Data} (O'Reilly
  Media)

\bibitem[{{Cowie} {et~al.}(1996){Cowie}, {Songaila}, {Hu}, \&
  {Cohen}}]{Cowie1996}
{Cowie}, L.~L., {Songaila}, A., {Hu}, E.~M., \& {Cohen}, J.~G. 1996, \aj, 112,
  839, \dodoi{10.1086/118058}

\bibitem[{{Dekel} \& {Silk}(1986)}]{Dekel1986}
{Dekel}, A., \& {Silk}, J. 1986, \apj, 303, 39, \dodoi{10.1086/164050}

\bibitem[{{Digby} {et~al.}(2019){Digby}, {Navarro}, {Fattahi}, {Simpson},
  {Oman}, {Gomez}, {Frenk}, {Grand}, \& {Pakmor}}]{Digby2019}
{Digby}, R., {Navarro}, J.~F., {Fattahi}, A., {et~al.} 2019, \mnras, 485, 5423,
  \dodoi{10.1093/mnras/stz745}

\bibitem[{{Dubois} {et~al.}(2016){Dubois}, {Peirani}, {Pichon}, {Devriendt},
  {Gavazzi}, {Welker}, \& {Volonteri}}]{Dubois2016}
{Dubois}, Y., {Peirani}, S., {Pichon}, C., {et~al.} 2016, \mnras, 463, 3948,
  \dodoi{10.1093/mnras/stw2265}

\bibitem[{{El-Badry} {et~al.}(2016){El-Badry}, {Wetzel}, {Geha}, {Hopkins},
  {Kere{\v{s}}}, {Chan}, \& {Faucher-Gigu{\`e}re}}]{ElBadry2016}
{El-Badry}, K., {Wetzel}, A., {Geha}, M., {et~al.} 2016, \apj, 820, 131,
  \dodoi{10.3847/0004-637X/820/2/131}

\bibitem[{{Escala} {et~al.}(2018){Escala}, {Wetzel}, {Kirby}, {Hopkins}, {Ma},
  {Wheeler}, {Kere{\v{s}}}, {Faucher-Gigu{\`e}re}, \& {Quataert}}]{Escala2018}
{Escala}, I., {Wetzel}, A., {Kirby}, E.~N., {et~al.} 2018, \mnras, 474, 2194,
  \dodoi{10.1093/mnras/stx2858}

\bibitem[{{Faucher-Gigu{\`e}re}(2020)}]{FaucherGiguere2020}
{Faucher-Gigu{\`e}re}, C.-A. 2020, \mnras, 493, 1614,
  \dodoi{10.1093/mnras/staa302}

\bibitem[{{Faucher-Gigu{\`e}re} {et~al.}(2009){Faucher-Gigu{\`e}re}, {Lidz},
  {Zaldarriaga}, \& {Hernquist}}]{FaucherGiguere2009}
{Faucher-Gigu{\`e}re}, C.-A., {Lidz}, A., {Zaldarriaga}, M., \& {Hernquist}, L.
  2009, \apj, 703, 1416, \dodoi{10.1088/0004-637X/703/2/1416}

\bibitem[{{Feldmann} {et~al.}(2019){Feldmann}, {Faucher-Gigu{\`e}re}, \&
  {Kere{\v{s}}}}]{Feldmann2019}
{Feldmann}, R., {Faucher-Gigu{\`e}re}, C.-A., \& {Kere{\v{s}}}, D. 2019, \apjl,
  871, L21, \dodoi{10.3847/2041-8213/aafe80}

\bibitem[{{Feldmann} {et~al.}(2023){Feldmann}, {Quataert},
  {Faucher-Gigu{\`e}re}, {Hopkins}, {{\c{C}}atmabacak}, {Kere{\v{s}}},
  {Bassini}, {Bernardini}, {Bullock}, {Cenci}, {Gensior}, {Liang}, {Moreno}, \&
  {Wetzel}}]{Feldmann2023}
{Feldmann}, R., {Quataert}, E., {Faucher-Gigu{\`e}re}, C.-A., {et~al.} 2023,
  \mnras, 522, 3831, \dodoi{10.1093/mnras/stad1205}

\bibitem[{{Fillingham} {et~al.}(2016){Fillingham}, {Cooper}, {Pace},
  {Boylan-Kolchin}, {Bullock}, {Garrison-Kimmel}, \&
  {Wheeler}}]{Fillingham2016}
{Fillingham}, S.~P., {Cooper}, M.~C., {Pace}, A.~B., {et~al.} 2016, \mnras,
  463, 1916, \dodoi{10.1093/mnras/stw2131}

\bibitem[{{Fillingham} {et~al.}(2015){Fillingham}, {Cooper}, {Wheeler},
  {Garrison-Kimmel}, {Boylan-Kolchin}, \& {Bullock}}]{Fillingham2015}
{Fillingham}, S.~P., {Cooper}, M.~C., {Wheeler}, C., {et~al.} 2015, \mnras,
  454, 2039, \dodoi{10.1093/mnras/stv2058}

\bibitem[{{Fofie} {et~al.}(2026){Fofie}, {Bullock}, {Wellons}, {Moreno},
  {Mercado}, \& {Wetzel}}]{Fofie2026}
{Fofie}, P., {Bullock}, J.~S., {Wellons}, S., {et~al.} 2026, arXiv e-prints,
  arXiv:2608.13718, \dodoi{10.48550/arXiv.2608.13718}

\bibitem[{{Fontanot} {et~al.}(2009){Fontanot}, {De Lucia}, {Monaco},
  {Somerville}, \& {Santini}}]{Fontanot2009}
{Fontanot}, F., {De Lucia}, G., {Monaco}, P., {Somerville}, R.~S., \&
  {Santini}, P. 2009, \mnras, 397, 1776,
  \dodoi{10.1111/j.1365-2966.2009.15058.x}

\bibitem[{{Gandhi} {et~al.}(2022){Gandhi}, {Wetzel}, {Hopkins}, {Shappee},
  {Wheeler}, \& {Faucher-Gigu{\`e}re}}]{Gandhi2022}
{Gandhi}, P.~J., {Wetzel}, A., {Hopkins}, P.~F., {et~al.} 2022, \mnras, 516,
  1941, \dodoi{10.1093/mnras/stac2228}

\bibitem[{{Gao} {et~al.}(2005){Gao}, {Springel}, \& {White}}]{Gao2005}
{Gao}, L., {Springel}, V., \& {White}, S. D.~M. 2005, \mnras, 363, L66,
  \dodoi{10.1111/j.1745-3933.2005.00084.x}

\bibitem[{{Garrison-Kimmel} {et~al.}(2019{\natexlab{a}}){Garrison-Kimmel},
  {Wetzel}, {Hopkins}, {Sanderson}, {El-Badry}, {Graus}, {Chan}, {Feldmann},
  {Boylan-Kolchin}, {Hayward}, {Bullock}, {Fitts}, {Samuel}, {Wheeler},
  {Kere{\v{s}}}, \& {Faucher-Gigu{\`e}re}}]{GarrisonKimmel2019b}
{Garrison-Kimmel}, S., {Wetzel}, A., {Hopkins}, P.~F., {et~al.}
  2019{\natexlab{a}}, \mnras, 489, 4574, \dodoi{10.1093/mnras/stz2507}

\bibitem[{{Garrison-Kimmel} {et~al.}(2019{\natexlab{b}}){Garrison-Kimmel},
  {Hopkins}, {Wetzel}, {Bullock}, {Boylan-Kolchin}, {Kere{\v{s}}},
  {Faucher-Gigu{\`e}re}, {El-Badry}, {Lamberts}, {Quataert}, \&
  {Sanderson}}]{GarrisonKimmel2019a}
{Garrison-Kimmel}, S., {Hopkins}, P.~F., {Wetzel}, A., {et~al.}
  2019{\natexlab{b}}, \mnras, 487, 1380, \dodoi{10.1093/mnras/stz1317}

\bibitem[{{Geng} {et~al.}(2026){Geng}, {Feldmann}, {Moreno}, {Shen}, \&
  {Bullock}}]{Geng2026}
{Geng}, H., {Feldmann}, R., {Moreno}, J., {Shen}, X., \& {Bullock}, J. 2026,
  arXiv e-prints, arXiv:2610.26969

\bibitem[{{Goto} {et~al.}(2003){Goto}, {Yamauchi}, {Fujita}, {Okamura},
  {Sekiguchi}, {Smail}, {Bernardi}, \& {Gomez}}]{Goto2003}
{Goto}, T., {Yamauchi}, C., {Fujita}, Y., {et~al.} 2003, \mnras, 346, 601,
  \dodoi{10.1046/j.1365-2966.2003.07114.x}

\bibitem[{{Graf} {et~al.}(2025{\natexlab{a}}){Graf}, {Wetzel}, {Bailin}, \&
  {Orr}}]{Graf2025b}
{Graf}, R.~L., {Wetzel}, A., {Bailin}, J., \& {Orr}, M.~E. 2025{\natexlab{a}},
  \apj, 991, 139, \dodoi{10.3847/1538-4357/adfa07}

\bibitem[{{Graf} {et~al.}(2025{\natexlab{b}}){Graf}, {Wetzel}, {Bellardini}, \&
  {Bailin}}]{Graf2025a}
{Graf}, R.~L., {Wetzel}, A., {Bellardini}, M.~A., \& {Bailin}, J.
  2025{\natexlab{b}}, \apj, 981, 47, \dodoi{10.3847/1538-4357/adacd7}

\bibitem[{{Grand} {et~al.}(2017){Grand}, {G{\'o}mez}, {Marinacci}, {Pakmor},
  {Springel}, {Campbell}, {Frenk}, {Jenkins}, \& {White}}]{Grand2017}
{Grand}, R. J.~J., {G{\'o}mez}, F.~A., {Marinacci}, F., {et~al.} 2017, \mnras,
  467, 179, \dodoi{10.1093/mnras/stx071}

\bibitem[{{Guglielmo} {et~al.}(2015){Guglielmo}, {Poggianti}, {Moretti},
  {Fritz}, {Calvi}, {Vulcani}, {Fasano}, \& {Paccagnella}}]{Guglielmo2015}
{Guglielmo}, V., {Poggianti}, B.~M., {Moretti}, A., {et~al.} 2015, \mnras, 450,
  2749, \dodoi{10.1093/mnras/stv757}

\bibitem[{{Hahn} \& {Abel}(2011)}]{Hahn2011}
{Hahn}, O., \& {Abel}, T. 2011, \mnras, 415, 2101,
  \dodoi{10.1111/j.1365-2966.2011.18820.x}

\bibitem[{{Harris} {et~al.}(2020){Harris}, {Millman}, {van der Walt},
  {Gommers}, {Virtanen}, {Cournapeau}, {Wieser}, {Taylor}, {Berg}, {Smith},
  {Kern}, {Picus}, {Hoyer}, {van Kerkwijk}, {Brett}, {Haldane}, {Fern{\'a}ndez
  del R{\'i}o}, {Wiebe}, {Peterson}, {G{\'e}rard-Marchant}, {Sheppard},
  {Reddy}, {Weckesser}, {Abbasi}, {Gohlke}, \& {Oliphant}}]{Harris2020}
{Harris}, C.~R., {Millman}, K.~J., {van der Walt}, S.~J., {et~al.} 2020,
  Nature, 585, 357, \dodoi{10.1038/s41586-020-2649-2}

\bibitem[{{Hopkins}(2015)}]{Hopkins2015}
{Hopkins}, P.~F. 2015, \mnras, 450, 53, \dodoi{10.1093/mnras/stv195}

\bibitem[{{Hopkins} {et~al.}(2018){Hopkins}, {Wetzel}, {Kere{\v{s}}},
  {Faucher-Gigu{\`e}re}, {Quataert}, {Boylan-Kolchin}, {Murray}, {Hayward},
  {Garrison-Kimmel}, {Hummels}, {Feldmann}, {Torrey}, {Ma},
  {Angl{\'e}s-Alc{\'a}zar}, {Su}, {Orr}, {Schmitz}, {Escala}, {Sanderson},
  {Grudi{\'c}}, {Hafen}, {Kim}, {Fitts}, {Bullock}, {Wheeler}, {Chan},
  {Elbert}, \& {Narayanan}}]{Hopkins2018}
{Hopkins}, P.~F., {Wetzel}, A., {Kere{\v{s}}}, D., {et~al.} 2018, \mnras, 480,
  800, \dodoi{10.1093/mnras/sty1690}

\bibitem[{{Hunter}(2007)}]{Hunter2007}
{Hunter}, J.~D. 2007, Computing in Science and Engineering, 9, 90,
  \dodoi{10.1109/MCSE.2007.55}

\bibitem[{{Joshi} {et~al.}(2021){Joshi}, {Pillepich}, {Nelson}, {Zinger},
  {Marinacci}, {Springel}, {Vogelsberger}, \& {Hernquist}}]{Joshi2021}
{Joshi}, G.~D., {Pillepich}, A., {Nelson}, D., {et~al.} 2021, \mnras, 508,
  1652, \dodoi{10.1093/mnras/stab2573}

\bibitem[{{Kado-Fong} {et~al.}(2020){Kado-Fong}, {Greene}, {Huang}, {Beaton},
  {Goulding}, \& {Komiyama}}]{KadoFong2020}
{Kado-Fong}, E., {Greene}, J.~E., {Huang}, S., {et~al.} 2020, \apj, 900, 163,
  \dodoi{10.3847/1538-4357/abacc2}

\bibitem[{{Kauffmann} {et~al.}(2004){Kauffmann}, {White}, {Heckman},
  {M{\'e}nard}, {Brinchmann}, {Charlot}, {Tremonti}, \&
  {Brinkmann}}]{Kauffmann2004}
{Kauffmann}, G., {White}, S. D.~M., {Heckman}, T.~M., {et~al.} 2004, \mnras,
  353, 713, \dodoi{10.1111/j.1365-2966.2004.08117.x}

\bibitem[{{Kazantzidis} {et~al.}(2011){Kazantzidis}, {{\L}okas}, {Callegari},
  {Mayer}, \& {Moustakas}}]{Kazantzidis2011}
{Kazantzidis}, S., {{\L}okas}, E.~L., {Callegari}, S., {Mayer}, L., \&
  {Moustakas}, L.~A. 2011, \apj, 726, 98, \dodoi{10.1088/0004-637X/726/2/98}

\bibitem[{{Klein} {et~al.}(2026){Klein}, {Wang}, {Xia}, {Bullock}, {Moreno},
  {Feldmann}, {Mercado}, {Faucher-Gigu{\`e}re}, {Stern}, {Sanchez}, {Hussein},
  \& {Park}}]{Klein2026}
{Klein}, C., {Wang}, J.~D., {Xia}, L.~Y., {et~al.} 2026, \apj, 998, 125,
  \dodoi{10.3847/1538-4357/ae31f4}

\bibitem[{{Knollmann} \& {Knebe}(2009)}]{Knollmann2009}
{Knollmann}, S.~R., \& {Knebe}, A. 2009, \apjs, 182, 608,
  \dodoi{10.1088/0067-0049/182/2/608}

\bibitem[{{Kroupa}(2002)}]{Kroupa2002}
{Kroupa}, P. 2002, Science, 295, 82, \dodoi{10.1126/science.1067524}

\bibitem[{{Leitherer} {et~al.}(1999){Leitherer}, {Schaerer}, {Goldader},
  {Delgado}, {Robert}, {Kune}, {de Mello}, {Devost}, \&
  {Heckman}}]{Leitherer1999}
{Leitherer}, C., {Schaerer}, D., {Goldader}, J.~D., {et~al.} 1999, \apjs, 123,
  3, \dodoi{10.1086/313233}

\bibitem[{{{\L}okas} {et~al.}(2012){{\L}okas}, {Majewski}, {Kazantzidis},
  {Mayer}, {Carlin}, {Nidever}, \& {Moustakas}}]{Lokas2012}
{{\L}okas}, E.~L., {Majewski}, S.~R., {Kazantzidis}, S., {et~al.} 2012, \apj,
  751, 61, \dodoi{10.1088/0004-637X/751/1/61}

\bibitem[{{{\L}okas} {et~al.}(2015){{\L}okas}, {Semczuk}, {Gajda}, \&
  {D'Onghia}}]{Lokas2015}
{{\L}okas}, E.~L., {Semczuk}, M., {Gajda}, G., \& {D'Onghia}, E. 2015, \apj,
  810, 100, \dodoi{10.1088/0004-637X/810/2/100}

\bibitem[{{Ludlow} {et~al.}(2021){Ludlow}, {Fall}, {Schaye}, \&
  {Obreschkow}}]{Ludlow2021}
{Ludlow}, A.~D., {Fall}, S.~M., {Schaye}, J., \& {Obreschkow}, D. 2021, \mnras,
  508, 5114, \dodoi{10.1093/mnras/stab2770}

\bibitem[{{Ludlow} {et~al.}(2023){Ludlow}, {Fall}, {Wilkinson}, {Schaye}, \&
  {Obreschkow}}]{Ludlow2023}
{Ludlow}, A.~D., {Fall}, S.~M., {Wilkinson}, M.~J., {Schaye}, J., \&
  {Obreschkow}, D. 2023, \mnras, 525, 5614, \dodoi{10.1093/mnras/stad2615}

\bibitem[{{Mayer} {et~al.}(2001){Mayer}, {Governato}, {Colpi}, {Moore},
  {Quinn}, {Wadsley}, {Stadel}, \& {Lake}}]{Mayer2001}
{Mayer}, L., {Governato}, F., {Colpi}, M., {et~al.} 2001, \apjl, 547, L123,
  \dodoi{10.1086/318898}

\bibitem[{{Mayer} {et~al.}(2006){Mayer}, {Mastropietro}, {Wadsley}, {Stadel},
  \& {Moore}}]{Mayer2006}
{Mayer}, L., {Mastropietro}, C., {Wadsley}, J., {Stadel}, J., \& {Moore}, B.
  2006, \mnras, 369, 1021, \dodoi{10.1111/j.1365-2966.2006.10403.x}

\bibitem[{{McCluskey} {et~al.}(2025){McCluskey}, {Wetzel}, {Loebman}, \&
  {Moreno}}]{McCluskey2025}
{McCluskey}, F., {Wetzel}, A., {Loebman}, S., \& {Moreno}, J. 2025, arXiv
  e-prints, arXiv:2506.11840

\bibitem[{{McConnachie}(2012)}]{McConnachie2012}
{McConnachie}, A.~W. 2012, \aj, 144, 4, \dodoi{10.1088/0004-6256/144/1/4}

\bibitem[{{McKinney}(2010)}]{McKinney2010}
{McKinney}, W. 2010, in Proceedings of the 9th Python in Science Conference,
  ed. S.~{van der Walt} \& J.~{Millman}, 56--61,
  \dodoi{10.25080/Majora-92bf1922-00a}

\bibitem[{{Mercado} {et~al.}(2025){Mercado}, {Moreno}, {Feldmann}, {Zeender},
  {Benavides}, {Piotrowska}, {Klein}, {Wheeler}, {Necib}, {Bullock}, \&
  {Hopkins}}]{Mercado2025}
{Mercado}, F.~J., {Moreno}, J., {Feldmann}, R., {et~al.} 2025, \apj, 983, 93,
  \dodoi{10.3847/1538-4357/adbf07}

\bibitem[{{Mercado} {et~al.}(2026){Mercado}, {Baxter}, {Rodriguez Wimberly},
  {Moreno}, {Wheeler}, {Gandhi}, {Wetzel}, {Feldmann}, {Tortora}, \&
  {Samuel}}]{Mercado2026}
{Mercado}, F.~J., {Baxter}, D.~C., {Rodriguez Wimberly}, M.~K., {et~al.} 2026,
  \apj, 1004, 217, \dodoi{10.3847/1538-4357/ae7360}

\bibitem[{{Mo} {et~al.}(2010){Mo}, {van den Bosch}, \& {White}}]{Mo2010}
{Mo}, H., {van den Bosch}, F.~C., \& {White}, S. 2010, {Galaxy Formation and
  Evolution} (Cambrige University Press)

\bibitem[{{Montero-Dorta} {et~al.}(2021){Montero-Dorta}, {Chaves-Montero},
  {Artale}, \& {Favole}}]{MonteroDorta2021}
{Montero-Dorta}, A.~D., {Chaves-Montero}, J., {Artale}, M.~C., \& {Favole}, G.
  2021, \mnras, 508, 940, \dodoi{10.1093/mnras/stab2556}

\bibitem[{{Moreno} {et~al.}(2022){Moreno}, {Danieli}, {Bullock}, {Feldmann},
  {Hopkins}, {{\c{c}}atmabacak}, {Gurvich}, {Lazar}, {Klein}, {Hummels},
  {Hafen}, {Mercado}, {Yu}, {Jiang}, {Wheeler}, {Wetzel},
  {Angl{\'e}s-Alc{\'a}zar}, {Boylan-Kolchin}, {Quataert},
  {Faucher-Gigu{\`e}re}, \& {Kere{\v{s}}}}]{Moreno2022}
{Moreno}, J., {Danieli}, S., {Bullock}, J.~S., {et~al.} 2022, Nature Astronomy,
  6, 496, \dodoi{10.1038/s41550-021-01598-4}

\bibitem[{{Moreno} {et~al.}(2025){Moreno}, {Wheeler}, {Mercado}, {Rodriguez
  Wimberly}, {Samuel}, {Gandhi}, {Cenci}, {Feldmann}, {Boylan-Kolchin},
  {Wetzel}, {Bullock}, \& {Hopkins}}]{Moreno2025}
{Moreno}, J., {Wheeler}, C., {Mercado}, F.~J., {et~al.} 2025, arXiv e-prints,
  arXiv:2511.03785, \dodoi{10.48550/arXiv.2511.03785}

\bibitem[{{Mowla} {et~al.}(2019){Mowla}, {van der Wel}, {van Dokkum}, \&
  {Miller}}]{Mowla2019}
{Mowla}, L., {van der Wel}, A., {van Dokkum}, P., \& {Miller}, T.~B. 2019,
  \apjl, 872, L13, \dodoi{10.3847/2041-8213/ab0379}

\bibitem[{{Myrtaj} {et~al.}(2026){Myrtaj}, {Bullock}, {Boylan-Kolchin},
  {Chandra}, {Faucher-Gigu{\`e}re}, {Feldmann}, {Gandhi}, {Mercado}, {Moreno},
  {Stern}, {Wetzel}, \& {Xia}}]{Myrtaj2026}
{Myrtaj}, O., {Bullock}, J.~S., {Boylan-Kolchin}, M., {et~al.} 2026, \mnras,
  553, \dodoi{10.1093/mnras/stag1902}

\bibitem[{{Naab} \& {Ostriker}(2017)}]{Naab2017}
{Naab}, T., \& {Ostriker}, J.~P. 2017, \araa, 55, 59,
  \dodoi{10.1146/annurev-astro-081913-040019}

\bibitem[{{Peng} {et~al.}(2010){Peng}, {Lilly}, {Kova{\v{c}}}, {Bolzonella},
  {Pozzetti}, {Renzini}, {Zamorani}, {Ilbert}, {Knobel}, {Iovino}, {Maier},
  {Cucciati}, {Tasca}, {Carollo}, {Silverman}, {Kampczyk}, {de Ravel},
  {Sanders}, {Scoville}, {Contini}, {Mainieri}, {Scodeggio}, {Kneib}, {Le
  F{\`e}vre}, {Bardelli}, {Bongiorno}, {Caputi}, {Coppa}, {de la Torre},
  {Franzetti}, {Garilli}, {Lamareille}, {Le Borgne}, {Le Brun}, {Mignoli},
  {Perez Montero}, {Pello}, {Ricciardelli}, {Tanaka}, {Tresse}, {Vergani},
  {Welikala}, {Zucca}, {Oesch}, {Abbas}, {Barnes}, {Bordoloi}, {Bottini},
  {Cappi}, {Cassata}, {Cimatti}, {Fumana}, {Hasinger}, {Koekemoer},
  {Leauthaud}, {Maccagni}, {Marinoni}, {McCracken}, {Memeo}, {Meneux}, {Nair},
  {Porciani}, {Presotto}, \& {Scaramella}}]{Peng2010}
{Peng}, Y.-j., {Lilly}, S.~J., {Kova{\v{c}}}, K., {et~al.} 2010, \apj, 721,
  193, \dodoi{10.1088/0004-637X/721/1/193}

\bibitem[{{P{\'e}rez-Mill{\'a}n} {et~al.}(2023){P{\'e}rez-Mill{\'a}n}, {Fritz},
  {Gonz{\'a}lez-L{\'o}pezlira}, {Moretti}, {Cervantes Sodi}, {Vulcani},
  {Gullieuszik}, {Bruzual}, {Charlot}, \& {Bettoni}}]{PerezMillan2023}
{P{\'e}rez-Mill{\'a}n}, D., {Fritz}, J., {Gonz{\'a}lez-L{\'o}pezlira}, R.~A.,
  {et~al.} 2023, \mnras, 521, 1292, \dodoi{10.1093/mnras/stad542}

\bibitem[{{Pillepich} {et~al.}(2018){Pillepich}, {Springel}, {Nelson}, {Genel},
  {Naiman}, {Pakmor}, {Hernquist}, {Torrey}, {Vogelsberger}, {Weinberger}, \&
  {Marinacci}}]{Pillepich2018b}
{Pillepich}, A., {Springel}, V., {Nelson}, D., {et~al.} 2018, \mnras, 473,
  4077, \dodoi{10.1093/mnras/stx2656}

\bibitem[{{Planck Collaboration} {et~al.}(2016){Planck Collaboration}, {Ade},
  {Aghanim}, {Arnaud}, {Ashdown}, {Aumont}, {Baccigalupi}, {Banday},
  {Barreiro}, {Bartlett}, {Bartolo}, {Battaner}, {Battye}, {Benabed},
  {Beno{\^\i}t}, {Benoit-L{\'e}vy}, {Bernard}, {Bersanelli}, {Bielewicz},
  {Bock}, {Bonaldi}, {Bonavera}, {Bond}, {Borrill}, {Bouchet}, {Boulanger},
  {Bucher}, {Burigana}, {Butler}, {Calabrese}, {Cardoso}, {Catalano},
  {Challinor}, {Chamballu}, {Chary}, {Chiang}, {Chluba}, {Christensen},
  {Church}, {Clements}, {Colombi}, {Colombo}, {Combet}, {Coulais}, {Crill},
  {Curto}, {Cuttaia}, {Danese}, {Davies}, {Davis}, {de Bernardis}, {de Rosa},
  {de Zotti}, {Delabrouille}, {D{\'e}sert}, {Di Valentino}, {Dickinson},
  {Diego}, {Dolag}, {Dole}, {Donzelli}, {Dor{\'e}}, {Douspis}, {Ducout},
  {Dunkley}, {Dupac}, {Efstathiou}, {Elsner}, {En{\ss}lin}, {Eriksen},
  {Farhang}, {Fergusson}, {Finelli}, {Forni}, {Frailis}, {Fraisse},
  {Franceschi}, {Frejsel}, {Galeotta}, {Galli}, {Ganga}, {Gauthier}, {Gerbino},
  {Ghosh}, {Giard}, {Giraud-H{\'e}raud}, {Giusarma}, {Gjerl{\o}w},
  {Gonz{\'a}lez-Nuevo}, {G{\'o}rski}, {Gratton}, {Gregorio}, {Gruppuso},
  {Gudmundsson}, {Hamann}, {Hansen}, {Hanson}, {Harrison}, {Helou},
  {Henrot-Versill{\'e}}, {Hern{\'a}ndez-Monteagudo}, {Herranz}, {Hildebrandt},
  {Hivon}, {Hobson}, {Holmes}, {Hornstrup}, {Hovest}, {Huang}, {Huffenberger},
  {Hurier}, {Jaffe}, {Jaffe}, {Jones}, {Juvela}, {Keih{\"a}nen}, {Keskitalo},
  {Kisner}, {Kneissl}, {Knoche}, {Knox}, {Kunz}, {Kurki-Suonio}, {Lagache},
  {L{\"a}hteenm{\"a}ki}, {Lamarre}, {Lasenby}, {Lattanzi}, {Lawrence}, {Leahy},
  {Leonardi}, {Lesgourgues}, {Levrier}, {Lewis}, {Liguori}, {Lilje},
  {Linden-V{\o}rnle}, {L{\'o}pez-Caniego}, {Lubin}, {Mac{\'\i}as-P{\'e}rez},
  {Maggio}, {Maino}, {Mandolesi}, {Mangilli}, {Marchini}, {Maris}, {Martin},
  {Martinelli}, {Mart{\'\i}nez-Gonz{\'a}lez}, {Masi}, {Matarrese}, {McGehee},
  {Meinhold}, {Melchiorri}, {Melin}, {Mendes}, {Mennella}, {Migliaccio},
  {Millea}, {Mitra}, {Miville-Desch{\^e}nes}, {Moneti}, {Montier}, {Morgante},
  {Mortlock}, {Moss}, {Munshi}, {Murphy}, {Naselsky}, {Nati}, {Natoli},
  {Netterfield}, {N{\o}rgaard-Nielsen}, {Noviello}, {Novikov}, {Novikov},
  {Oxborrow}, {Paci}, {Pagano}, {Pajot}, {Paladini}, {Paoletti}, {Partridge},
  {Pasian}, {Patanchon}, {Pearson}, {Perdereau}, {Perotto}, {Perrotta},
  {Pettorino}, {Piacentini}, {Piat}, {Pierpaoli}, {Pietrobon}, {Plaszczynski},
  {Pointecouteau}, {Polenta}, {Popa}, {Pratt}, {Pr{\'e}zeau}, {Prunet},
  {Puget}, {Rachen}, {Reach}, {Rebolo}, {Reinecke}, {Remazeilles}, {Renault},
  {Renzi}, {Ristorcelli}, {Rocha}, {Rosset}, {Rossetti}, {Roudier},
  {Rouill{\'e} d'Orfeuil}, {Rowan-Robinson}, {Rubi{\~n}o-Mart{\'\i}n},
  {Rusholme}, {Said}, {Salvatelli}, {Salvati}, {Sandri}, {Santos},
  {Savelainen}, {Savini}, {Scott}, {Seiffert}, {Serra}, {Shellard}, {Spencer},
  {Spinelli}, {Stolyarov}, {Stompor}, {Sudiwala}, {Sunyaev}, {Sutton},
  {Suur-Uski}, {Sygnet}, {Tauber}, {Terenzi}, {Toffolatti}, {Tomasi},
  {Tristram}, {Trombetti}, {Tucci}, {Tuovinen}, {T{\"u}rler}, {Umana},
  {Valenziano}, {Valiviita}, {Van Tent}, {Vielva}, {Villa}, {Wade}, {Wandelt},
  {Wehus}, {White}, {White}, {Wilkinson}, {Yvon}, {Zacchei}, \&
  {Zonca}}]{PlanckCollaboration2016}
{Planck Collaboration}, {Ade}, P.~A.~R., {Aghanim}, N., {et~al.} 2016, \aap,
  594, A13, \dodoi{10.1051/0004-6361/201525830}

\bibitem[{{Power} {et~al.}(2003){Power}, {Navarro}, {Jenkins}, {Frenk},
  {White}, {Springel}, {Stadel}, \& {Quinn}}]{Power2003}
{Power}, C., {Navarro}, J.~F., {Jenkins}, A., {et~al.} 2003, \mnras, 338, 14,
  \dodoi{10.1046/j.1365-8711.2003.05925.x}

\bibitem[{{Samuel} {et~al.}(2022){Samuel}, {Wetzel}, {Santistevan}, {Tollerud},
  {Moreno}, {Boylan-Kolchin}, {Bailin}, \& {Pardasani}}]{Samuel2022}
{Samuel}, J., {Wetzel}, A., {Santistevan}, I., {et~al.} 2022, \mnras, 514,
  5276, \dodoi{10.1093/mnras/stac1706}

\bibitem[{{Samuel} {et~al.}(2020){Samuel}, {Wetzel}, {Tollerud},
  {Garrison-Kimmel}, {Loebman}, {El-Badry}, {Hopkins}, {Boylan-Kolchin},
  {Faucher-Gigu{\`e}re}, {Bullock}, {Benincasa}, \& {Bailin}}]{Samuel2020}
{Samuel}, J., {Wetzel}, A., {Tollerud}, E., {et~al.} 2020, \mnras, 491, 1471,
  \dodoi{10.1093/mnras/stz3054}

\bibitem[{{Santistevan} {et~al.}(2020){Santistevan}, {Wetzel}, {El-Badry},
  {Bland-Hawthorn}, {Boylan-Kolchin}, {Bailin}, {Faucher-Gigu{\`e}re}, \&
  {Benincasa}}]{Santistevan2020}
{Santistevan}, I.~B., {Wetzel}, A., {El-Badry}, K., {et~al.} 2020, \mnras, 497,
  747, \dodoi{10.1093/mnras/staa1923}

\bibitem[{{Santistevan} {et~al.}(2023){Santistevan}, {Wetzel}, {Tollerud},
  {Sanderson}, \& {Samuel}}]{Santistevan2023}
{Santistevan}, I.~B., {Wetzel}, A., {Tollerud}, E., {Sanderson}, R.~E., \&
  {Samuel}, J. 2023, \mnras, 518, 1427, \dodoi{10.1093/mnras/stac3100}

\bibitem[{{Sawala} {et~al.}(2016){Sawala}, {Frenk}, {Fattahi}, {Navarro},
  {Bower}, {Crain}, {Dalla Vecchia}, {Furlong}, {Helly}, {Jenkins}, {Oman},
  {Schaller}, {Schaye}, {Theuns}, {Trayford}, \& {White}}]{Sawala2016}
{Sawala}, T., {Frenk}, C.~S., {Fattahi}, A., {et~al.} 2016, \mnras, 457, 1931,
  \dodoi{10.1093/mnras/stw145}

\bibitem[{{Schaye} {et~al.}(2015){Schaye}, {Crain}, {Bower}, {Furlong},
  {Schaller}, {Theuns}, {Dalla Vecchia}, {Frenk}, {McCarthy}, {Helly},
  {Jenkins}, {Rosas-Guevara}, {White}, {Baes}, {Booth}, {Camps}, {Navarro},
  {Qu}, {Rahmati}, {Sawala}, {Thomas}, \& {Trayford}}]{Schaye2015}
{Schaye}, J., {Crain}, R.~A., {Bower}, R.~G., {et~al.} 2015, \mnras, 446, 521,
  \dodoi{10.1093/mnras/stu2058}

\bibitem[{{Simpson} {et~al.}(2018){Simpson}, {Grand}, {G{\'o}mez}, {Marinacci},
  {Pakmor}, {Springel}, {Campbell}, \& {Frenk}}]{Simpson2018}
{Simpson}, C.~M., {Grand}, R. J.~J., {G{\'o}mez}, F.~A., {et~al.} 2018, \mnras,
  478, 548, \dodoi{10.1093/mnras/sty774}

\bibitem[{{Somerville} \& {Dav{\'e}}(2015)}]{Somerville2015}
{Somerville}, R.~S., \& {Dav{\'e}}, R. 2015, \araa, 53, 51,
  \dodoi{10.1146/annurev-astro-082812-140951}

\bibitem[{{Taylor} \& {Kobayashi}(2015)}]{Taylor2015}
{Taylor}, P., \& {Kobayashi}, C. 2015, \mnras, 448, 1835,
  \dodoi{10.1093/mnras/stv139}

\bibitem[{{Thomas} {et~al.}(2005){Thomas}, {Maraston}, {Bender}, \& {Mendes de
  Oliveira}}]{Thomas2005}
{Thomas}, D., {Maraston}, C., {Bender}, R., \& {Mendes de Oliveira}, C. 2005,
  \apj, 621, 673, \dodoi{10.1086/426932}

\bibitem[{{van der Wel}(2008)}]{vanderWel2008}
{van der Wel}, A. 2008, \apjl, 675, L13, \dodoi{10.1086/529432}

\bibitem[{{van der Wel} {et~al.}(2014){van der Wel}, {Chang}, {Bell}, {Holden},
  {Ferguson}, {Giavalisco}, {Rix}, {Skelton}, {Whitaker}, {Momcheva},
  {Brammer}, {Kassin}, {Martig}, {Dekel}, {Ceverino}, {Koo}, {Mozena}, {van
  Dokkum}, {Franx}, {Faber}, \& {Primack}}]{vanderWel2014b}
{van der Wel}, A., {Chang}, Y.-Y., {Bell}, E.~F., {et~al.} 2014, \apjl, 792,
  L6, \dodoi{10.1088/2041-8205/792/1/L6}

\bibitem[{{Verley} {et~al.}(2007){Verley}, {Leon}, {Verdes-Montenegro},
  {Combes}, {Sabater}, {Sulentic}, {Bergond}, {Espada}, {Garc{\'\i}a},
  {Lisenfeld}, \& {Odewahn}}]{Verley2007}
{Verley}, S., {Leon}, S., {Verdes-Montenegro}, L., {et~al.} 2007, \aap, 472,
  121, \dodoi{10.1051/0004-6361:20077481}

\bibitem[{{Virtanen} {et~al.}(2020){Virtanen}, {Gommers}, {Oliphant},
  {Haberland}, {Reddy}, {Cournapeau}, {Burovski}, {Peterson}, {Weckesser},
  {Bright}, {van der Walt}, {Brett}, {Wilson}, {Millman}, {Mayorov}, {Nelson},
  {Jones}, {Kern}, {Larson}, {Carey}, {Polat}, {Feng}, {Moore}, {VanderPlas},
  {Laxalde}, {Perktold}, {Cimrman}, {Henriksen}, {Quintero}, {Harris},
  {Archibald}, {Ribeiro}, {Pedregosa}, {van Mulbregt}, \& {SciPy 1. 0
  Contributors}}]{Virtanen2020}
{Virtanen}, P., {Gommers}, R., {Oliphant}, T.~E., {et~al.} 2020, Nature
  Methods, 17, 261, \dodoi{10.1038/s41592-019-0686-2}

\bibitem[{{Wang} {et~al.}(2024){Wang}, {Nadler}, {Mao}, {Wechsler}, {Abel},
  {Behroozi}, {Geha}, {Asali}, {de los Reyes}, {Kado-Fong}, {Kallivayalil},
  {Tollerud}, {Weiner}, \& {Wu}}]{Wang2024}
{Wang}, Y., {Nadler}, E.~O., {Mao}, Y.-Y., {et~al.} 2024, \apj, 976, 119,
  \dodoi{10.3847/1538-4357/ad7f4c}

\bibitem[{{Wechsler} {et~al.}(2006){Wechsler}, {Zentner}, {Bullock},
  {Kravtsov}, \& {Allgood}}]{Wechsler2006}
{Wechsler}, R.~H., {Zentner}, A.~R., {Bullock}, J.~S., {Kravtsov}, A.~V., \&
  {Allgood}, B. 2006, \apj, 652, 71, \dodoi{10.1086/507120}

\bibitem[{{Weisz} {et~al.}(2014){Weisz}, {Dolphin}, {Skillman}, {Holtzman},
  {Gilbert}, {Dalcanton}, \& {Williams}}]{Weisz2014}
{Weisz}, D.~R., {Dolphin}, A.~E., {Skillman}, E.~D., {et~al.} 2014, \apj, 789,
  147, \dodoi{10.1088/0004-637X/789/2/147}

\bibitem[{{Weisz} {et~al.}(2011){Weisz}, {Dalcanton}, {Williams}, {Gilbert},
  {Skillman}, {Seth}, {Dolphin}, {McQuinn}, {Gogarten}, {Holtzman}, {Rosema},
  {Cole}, {Karachentsev}, \& {Zaritsky}}]{Weisz2011}
{Weisz}, D.~R., {Dalcanton}, J.~J., {Williams}, B.~F., {et~al.} 2011, \apj,
  739, 5, \dodoi{10.1088/0004-637X/739/1/5}

\bibitem[{{Wetzel} {et~al.}(2023){Wetzel}, {Hayward}, {Sanderson}, {Ma},
  {Angl{\'e}s-Alc{\'a}zar}, {Feldmann}, {Chan}, {El-Badry}, {Wheeler},
  {Garrison-Kimmel}, {Nikakhtar}, {Panithanpaisal}, {Arora}, {Gurvich},
  {Samuel}, {Sameie}, {Pandya}, {Hafen}, {Hummels}, {Loebman},
  {Boylan-Kolchin}, {Bullock}, {Faucher-Gigu{\`e}re}, {Kere{\v{s}}},
  {Quataert}, \& {Hopkins}}]{Wetzel2023}
{Wetzel}, A., {Hayward}, C.~C., {Sanderson}, R.~E., {et~al.} 2023, \apjs, 265,
  44, \dodoi{10.3847/1538-4365/acb99a}

\bibitem[{{Wetzel} {et~al.}(2025){Wetzel}, {Samuel}, {Gandhi}, {Ponnada}, {Su},
  {Arora}, {Angles-Alcazar}, {Hayward}, {Sanderson}, {Feldmann}, {Cochrane},
  {Nikakhtar}, {Panithanpaisal}, {Benavides}, {Pandya}, {Grudic}, {Hummels},
  {Gurvich}, {Hafen}, {Ma}, {Garrison-Kimmel}, {Sameie}, {Chan}, {El-Badry},
  {Necib}, {Loebman}, {Wellons}, {Robles}, {Wheeler}, {Moreno}, {Stern},
  {Boylan-Kolchin}, {Bullock}, {Faucher-Giguere}, {Keres}, {Quataert}, \&
  {Hopkins}}]{Wetzel2025}
{Wetzel}, A., {Samuel}, J., {Gandhi}, P.~J., {et~al.} 2025, arXiv e-prints,
  arXiv:2508.06608, \dodoi{10.48550/arXiv.2508.06608}

\bibitem[{{Wetzel} {et~al.}(2016){Wetzel}, {Hopkins}, {Kim},
  {Faucher-Gigu{\`e}re}, {Kere{\v{s}}}, \& {Quataert}}]{Wetzel2016}
{Wetzel}, A.~R., {Hopkins}, P.~F., {Kim}, J.-h., {et~al.} 2016, \apjl, 827,
  L23, \dodoi{10.3847/2041-8205/827/2/L23}

\bibitem[{{Xia} {et~al.}(2026){Xia}, {Klein}, {Wang}, {Bullock},
  {Boylan-Kolchin}, {Caudillo}, {Moreno}, {Mercado}, \& {Feldmann}}]{Xia2026}
{Xia}, L.~Y., {Klein}, C., {Wang}, J.~D., {et~al.} 2026, \mnras, 549, stag964,
  \dodoi{10.1093/mnras/stag964}

\bibitem[{{Yu} {et~al.}(2023){Yu}, {Bullock}, {Gurvich}, {Hafen},
  {et~al.}}]{Yu2023}
{Yu}, S., {Bullock}, J.~S., {Gurvich}, A.~B., {Hafen}, Z., {et~al.} 2023,
  \mnras, 523, 6220, \dodoi{10.1093/mnras/stad1806}

\bibitem[{{Yu} {et~al.}(2021){Yu}, {Bullock}, {Klein}, {Stern}, {Wetzel}, {Ma},
  {Moreno}, {et~al.}}]{Yu2021}
{Yu}, S., {Bullock}, J.~S., {Klein}, C., {et~al.} 2021, \mnras, 505, 889,
  \dodoi{10.1093/mnras/stab1339}

\end{thebibliography}
\bibliographystyle{aasjournal}

\end{document}